\documentclass{ieeeaccess}
\usepackage{cite}
\usepackage{amsmath,amssymb,amsfonts}
\usepackage{algorithm,algorithmic}
\usepackage{graphicx}
\usepackage{textcomp}
\usepackage{comment}
\usepackage{bm}
\usepackage{multirow}
\usepackage{subfigure}
\usepackage{color}
\usepackage{array}

\makeatletter
\AtBeginDocument{\DeclareMathVersion{bold}
\SetSymbolFont{operators}{bold}{T1}{times}{b}{n}
\SetSymbolFont{NewLetters}{bold}{T1}{times}{b}{it}
\SetMathAlphabet{\mathrm}{bold}{T1}{times}{b}{n}
\SetMathAlphabet{\mathit}{bold}{T1}{times}{b}{it}
\SetMathAlphabet{\mathbf}{bold}{T1}{times}{b}{n}
\SetMathAlphabet{\mathtt}{bold}{OT1}{pcr}{b}{n}
\SetSymbolFont{symbols}{bold}{OMS}{cmsy}{b}{n}
\renewcommand\boldmath{\@nomath\boldmath\mathversion{bold}}}
\makeatother

\def\BibTeX{{\rm B\kern-.05em{\sc i\kern-.025em b}\kern-.08em
    T\kern-.1667em\lower.7ex\hbox{E}\kern-.125emX}}

\newcommand{\argmin}{\mathop{\rm arg~min}\limits}

\newcommand{\narrowminipage}[1]{%
\begin{minipage}[t]{\linewidth}
\baselineskip=0.85\baselineskip
#1
\end{minipage}
}

\graphicspath{{./imgs/}}

\newcommand{\jtextd}[1]{{}}

\begin{document}
\history{Date of publication xxxx 00, 0000, date of current version xxxx 00, 0000.}
\doi{10.1109/ACCESS.2024.0429000}

\title{Rectifying Adversarial Examples Using Their Vulnerabilities}
\author{
\uppercase{FUMIYA MORIMOTO}\authorrefmark{1}, 
\uppercase{RYUTO MORITA}\authorrefmark{1}, 
\uppercase{SATOSHI ONO} \authorrefmark{1} 
\IEEEmembership{Member, IEEE}
}

\address[1]{Department of Information Science and Biomedical Engineering,
Graduate School of Science and Engineering, Kagoshima University\\
1-21-40, Korimoto, Kagoshima, 890-0065 Japan \\}

\tfootnote{%
This work was supported in part by JSPS KAKENHI Grant Number JP
22K12196 and JST A-STEP JPMJTM20T0.
}

\markboth
{Morimoto \headeretal: Rectifying Adversarial Examples Using Their Vulnerabilities}
{Morimoto \headeretal: Rectifying Adversarial Examples Using Their Vulnerabilities}

\corresp{Corresponding author: SATOSHI ONO (e-mail: ono@ibe.kagoshima-u.ac.jp).}

\begin{abstract}
Deep neural network-based classifiers
are prone to errors when processing adversarial examples (AEs).
AEs are
minimally perturbed input data undetectable to humans posing 
significant
risks to
security-dependent applications.
Hence,  
extensive
research has been undertaken to develop defense mechanisms that mitigate their threats.
Most existing methods primarily focus on discriminating AEs based on
the input sample
features,
emphasizing AE detection without addressing
the correct sample categorization before an attack.
While some tasks may only require mere rejection on detected AEs, others
necessitate identifying the correct original input category
such as traffic sign recognition in autonomous driving.
The objective of this study is to propose
a method for rectifying AEs
to estimate the correct labels of their original inputs.
Our method is based on re-attacking AEs to move them beyond the decision boundary
for accurate label prediction, effectively addressing 
the issue of rectifying
minimally perceptible AEs created using white-box attack methods.
However,
challenge remains
with respect to effectively rectifying
AEs produced by black-box attacks at a distance from the
boundary, or those misclassified into low-confidence categories by 
targeted attacks.
By adopting 
a straightforward approach
of
only considering AEs
as inputs, the proposed method 
can address 
diverse attacks while avoiding the requirement of
parameter adjustments or preliminary training.
Results 
demonstrate that the proposed method 
exhibits 
consistent
performance
in rectifying AEs generated via various attack methods,
including targeted and black-box attacks.
Moreover, it outperforms conventional rectification and input
transformation methods in terms of stability against various attacks.

\end{abstract}

\begin{keywords}
Deep neural network,
Adversarial example, 
Adversarial defense, 
Artificial intelligence security,
Label correction
\end{keywords}

\titlepgskip=-21pt

\maketitle

\section{Introduction}
\label{sec:introduction}

Recent studies have shown that deep neural network (DNN)-based classifiers are 
susceptible
to misrecognizing adversarial examples (AEs), which 
are small and specially perturbed input data, 
imperceptible
to humans%
~\cite{szegedy2013intriguing}. 
This vulnerability poses severe problems in security-critical 
tasks such as traffic sign recognition in autonomous driving%
~\cite{duan2021adversarial, sun2020counteracting, 
gnanasambandam2021optical, sato2024invisible} and 
image-based personal authentication~\cite{mirsky2021creation, 
tariq2022real, qin2021vulnerabilities}.
Owing to the possible exploitation of AEs
in real-world applications, 
addressing
their vulnerability 
is critical
to ensure the safety and
security of applied systems.
Risks associated with directly 
integrating
DNNs into various systems have prompted
research into DNNs 
for the development of methods that 
protect against AEs.

For instance, input transformation
~\cite{xie2017mitigating,dziugaite2016study} is an
approach
that aims
to reduce the influence of AEs
through preprocessing such as image transformation.
However, 
because
the same transformation is
applied to all inputs, benign samples are equally distorted by the
transformation, reducing the classification accuracy.
Additionally, this approach requires preprocessing method development
depending on the DNN input data types such as images and audio, 
as well as task-specific parameter adjustments.

\begin{figure}[t]
  \centering
    \subfigure[AE Detector only]{\includegraphics[width=4.0cm]{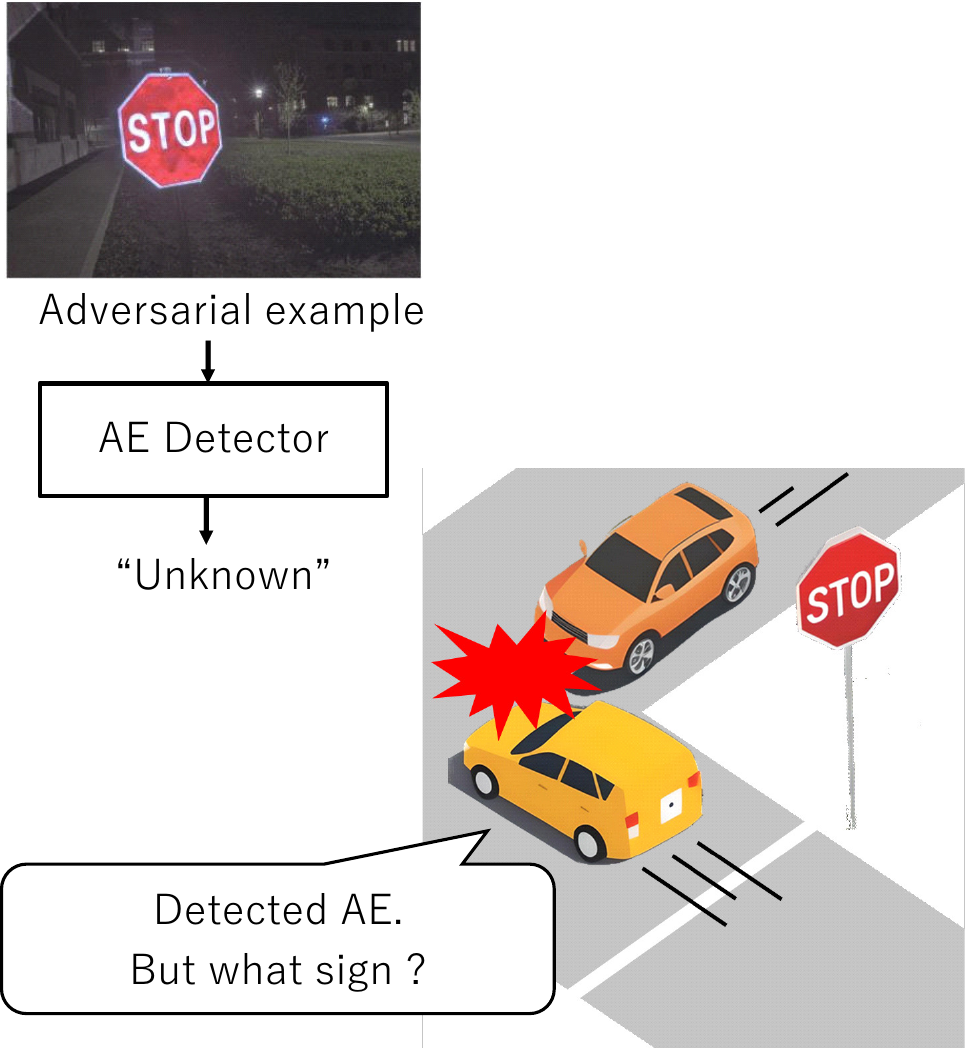}}
    ~
    \subfigure[Detector with proposed rectifier]{\includegraphics[width=4.0cm]{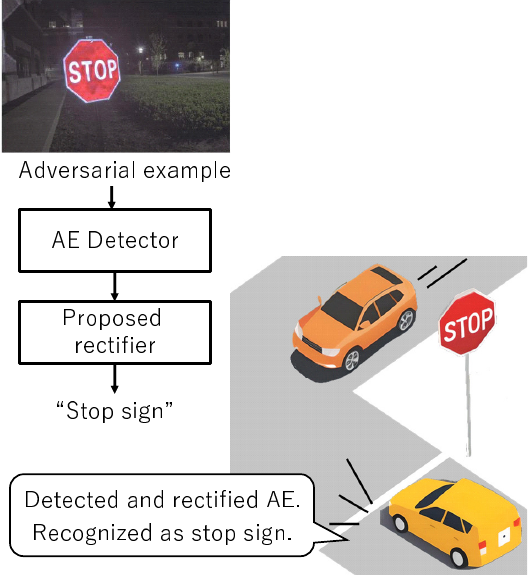}}
    \caption{Use case of our proposed method for road sign recognition in autonomous driving~\cite{gnanasambandam2021optical}.}
    \label{fig:overview}
\end{figure}

Meanwhile, detection methods~\cite{zhao2021attack} that discriminate
AEs based on input features and
maintain
recognition accuracy for
benign samples have been proposed.
However, they 
simply detect and discard 
AEs without recognizing the
correct category of pre-attack images. 
Although simply rejecting inputs detected as AEs may be sufficient for
numerous tasks, it 
becomes a critical issue in tasks requiring input
recognition before an attack.
For instance, in the case of a stop sign being attacked to confuse
autonomous cars, 
detecting
and discarding the attack as an AE are
insufficient as the car will not stop at the
correct locations
(Figure~\ref{fig:overview}(a)).
Furthermore, stopping the car upon detecting an AE is equally
hazardous
as the car may stop where it should not, for example, if a
speed limit sign is attacked.
Therefore, some postprocessing is required instead of rejecting
detected AEs so that they can be used to recognize a stop sign.

The task of correct classification label estimation from AE (i.e., the
label of an input sample before being attacked) is defined as
rectification~\cite{kao2022rectifying}.
As demonstrated in Figure~\ref{fig:overview}(b) with the example of
autonomous driving, combining rectification with an AE detector allows
an autonomous vehicle to stop properly at the stop sign, even if the
sign has been tampered with.
In addition to such critical tasks, detecting AEs and accurately
identifying their correct labels are advantageous for general
classification problems.
AE rectification has gained increased significance, as evidenced by
extensive research on input transformation and some studies on
detectors incorporating label
correction~\cite{wang2022detecting,yang2023best}.

Therefore, this study aims to propose a rectification method that
infers correct labels from AEs.
The proposed method focuses on the fragileness of AEs and 
re-attacks them to correct misclassification results so that they are
appropriately categorized to their original inputs.
Recent attack methods can generate minimally 
perturbed
 AEs that are
scarcely perceptible, depending on the characteristics of target DNN
models.
This means that generated AEs are located near classification
boundaries in the feature space, implying a strong probability of
changing the classification results when perturbations are added to
them.
Small perturbations to AEs that modify their classification results
are called the vulnerabilities of AEs~\cite{zhao2021attack}.
Given this fragility, re-attacking AEs using the proposed method can
effectively align misclassified results with the correct labels.

As defenders can usually access the internal information of the models
they defend, our approach utilizes a white-box attack method to
re-attack AEs.
By calculating gradients that reduce the confidence of misclassified
categories, the method efficiently re-attacks AEs to correct their
category.
Because our method
assumes that all inputs are AEs, it facilitates unrestricted re-attacks on
AEs, enabling continuous adjustments until the classification result
changes.
This alleviates the need for domain-specific pretraining, which
substantially benefits the proposed method.

The primary aim of this study is to examine the feasibility of
rectifying AEs produced by black-box attacks
to their correct labels through re-attacks.
The practical application of DNNs has accelerated in recent years,
leading to the availability of various image and audio processing
services as APIs.
Similarly, the number of commercial artificial intelligence (AI)
systems offered as cloud services, exemplified by generative AIs, is
rapidly increasing.
Many of these AI systems employ DNN models whose source code and
internal architecture information remain undisclosed, which
necessitates the use of black-box attack techniques when 
conducting
adversarial
attacks.
Consequently, the demand for methods to rectify AEs subjected to
black-box attacks is growing.

While it is anticipated that the proposed method will alter the
classification results of AEs by re-attacking them with gradient-based
perturbations, elucidating on means by which this method accurately
estimates the original input classes is imperative.
For instance, if the same attack method is employed to generate and
rectify an AE, our method is expected to readily discern the correct
class label of the original input from its AE.
However, if re-attacked using a different attack scheme than the one
initially employed, the method cannot always ascertain the original
input class.
Furthermore, substantial challenges are expected when attempting to
rectify AEs produced using black-box attacks devoid of gradient
information from a target DNN.

Another objective of this study is to examine the feasibility of
rectifying AEs generated through targeted attacks that misclassify the
original category as a category with low confidence using the proposed
method.
Adversarial attacks are typically classified into two categories:
untargeted attacks (attacks that misclassify an input to a label
different from the original classification result) and targeted
attacks (attacks that intentionally misclassify an input to a chosen
target label).
While untargeted attacks aim to reduce the confidence level of correct
predictions, targeted attacks seek to increase the confidence level of
the targeted predictions.
As the confidence of the category induced by the targeted attack
decreases, perturbations in AE increase, complicating accurate
category correction.

We validate the feasibility of the stable rectification of AEs back to
their correct labels using the proposed method, independent of
data types and defense models, through experiments across image- and
speech-recognition tasks 
involving up to seven attack methods.

The contribution of this study summarizes as follows:

\begin{itemize}

\item {\bf Training-free data- and detector-agnostic rectifier:}
By designing to operate independently from AE detectors, the proposed
method has advantages in terms of versatility, flexibility, and
efficiency.
Unlike input transformation preprocessing such as image smoothing,
which requires specific designs, implementations, and adjustments for
each type of input data and task, our method can be applied
universally, independent from the input data type.
Additionally, as AE detection is performed by the detector, the
proposed rectifier does not require prior trainings or adjustments.
This is because the rectifier is processed under the assumption that
the input is an AE, allowing re-attacks without presetting the
intensity or number of iterations; it only needs to continue until the
recognized label of the model changes.
Furthermore, our rectifier is seamlessly integrable with any detector,
enhancing the ease of developing new detectors and applications across
existing systems specialized for certain tasks or data types.

\item {\bf Experimental verification of the proposed method's robustness: }
Initially, a conceptual analysis of the method's effectiveness
against black-box attacks including score- and decision-based attacks
as well as targeted attacks is performed.
Then, the method's ability to address these challenges is empirically
demonstrated.
Additionally, the effectiveness of the proposed method against
aforementioned attacks is compared with an existing rectifier that
operates independently from detectors~\cite{kao2022rectifying}.
This previous method employs explainable artificial intelligence (XAI)
techniques to eliminate focus areas, achieving rectification and
showing superior performance than input transformation approaches.
The output comparison reveals the superior performance of the proposed
method, highlighting its advantages over input transformation methods
as well. 
Finally, the applicability of our method to audio and image modalities
is experimentally validated.

\end{itemize}

The subsequent sections of this paper are structured as follows: 
Section \ref{sec:Relatedwork}
reviews previous studies on adversarial attack and defense.
Section \ref{sec:Methodology} elucidates the fundamental concepts of the 
proposed method based on re-attacks to rectify AEs.
Section \ref{sec:Evaluation} 
describes the experimental settings and results, evaluating the
method’s effectiveness across various attacks and datasets in
image and audio modalities.
It elaborates on the comparison of our method with existing methods, 
presenting the method's
resilience
against targeted attacks and interaction with the detector.
Finally, Section \ref{sec:Conclusion} concludes the paper and
presents avenues for future research.

\section{Related work} 
\label{sec:Relatedwork}
\begin{figure}[t]
  \centering
  \includegraphics[width=8.0cm]{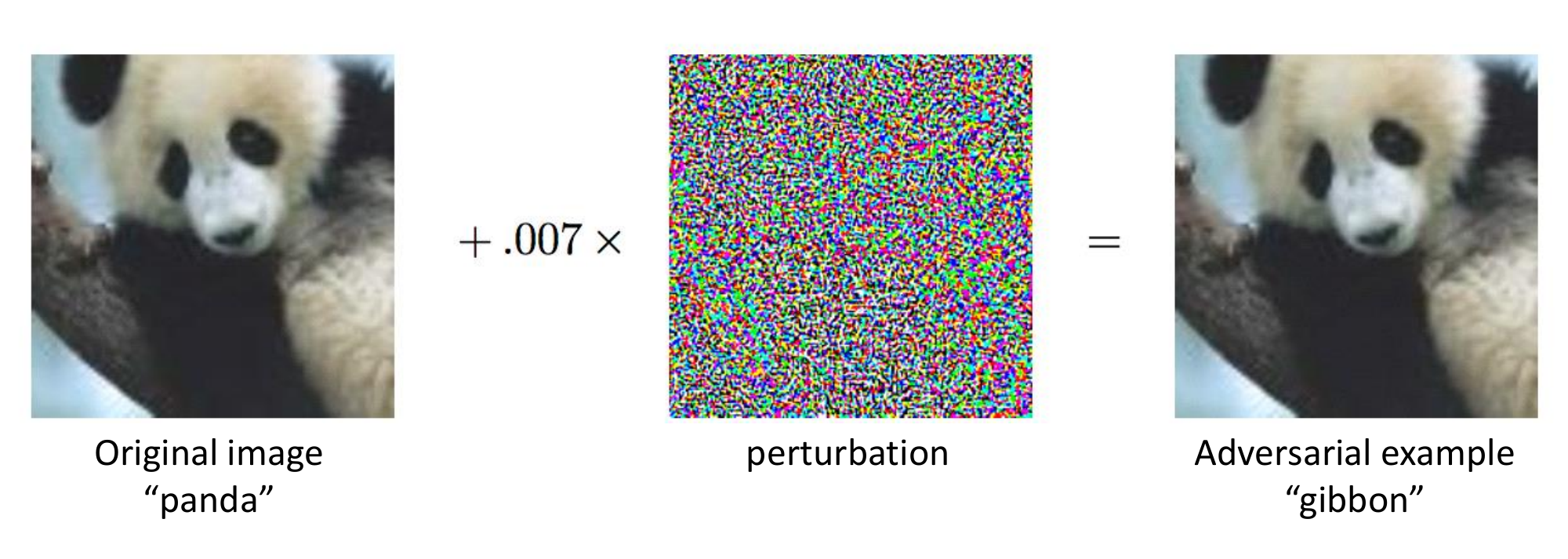}
  \caption{Example AE illustrated by Goodfellow et al.~\cite{goodfellow2014explaining}.}
  \label{fig:AE_exam}
\end{figure}

\subsection{Preliminaries}

In adversarial attacks, AEs are intentionally generated by an
adversary, akin to the example illustrated by Goodfellow et al
(Figure~\ref{fig:AE_exam})~\cite{goodfellow2014explaining}.
Here, ${\bm x'}$ is an AE generated by adding a small perturbation 
${\bm \delta}$ to an input ${\bm x}$, defined 
by the following equation,
\begin{displaymath}
  {\bm x'}={\bm x}+{\bm \delta}, \;\;\; ~ \; s.t.\;\; C({\bm x'}) \neq C({\bm x})
\end{displaymath}
where $C(\cdot)$ represents the classification result of a classifier. 
Perturbation ${\bm \delta}$ is defined as $L_p$ norm 
below $\varepsilon$ (${\|{\bm \delta}\|}_{p} < 
\varepsilon$).

We identify two types of adversary knowledge:
white-box and black-box. 
Under white-box scenarios, the adversary possesses the complete knowledge 
of the gradients and parameters of a target model. 
By contrast, under black-box scenarios, 
the adversary lacks knowledge regarding the model and cannot obtain 
various levels of its internal information.
White-box attacks, such as gradient-based attacks
~\cite{goodfellow2014explaining,kurakin2018adversarial,madry2017towards,moosavi2016deepfool,carlini2017towards,papernot2016limitations}, which utilize the model's gradient information, are potent.
However, under black-box scenarios, two types of attacks are possible:
score-based attacks~\cite{narodytska2016simple} and decision-based
attacks ~\cite{brendel2017decision, chen2020hopskipjumpattack}.
The former utilizes predictions and their confidence levels, while the
latter is based solely on predictions.

Generally, adversaries aim untargeted attacks, which misclassify input
samples to labels different from their original classification results
(untargeted attack scenario) or to a certain predetermined label
(targeted attack scenario).
Usually, untargeted attacks are more feasible than targeted ones. 
This is because
untargeted attacks decrease the confidence of correct predictions,
whereas targeted attacks aim to increase the confidence of targeted
predictions.

\subsection{Adversarial attack}
\subsubsection{Gradient-based attack}

Goodfellow et al. introduced the fast gradient sign method (FGSM)
~\cite{goodfellow2014explaining}, which generates AEs based on the
gradients of models.
It executes a one-step attack without iterations
for increasing
the gradient loss by one step along the gradient.
Kurakin et al. enhanced the attack performance of FGSM by
proposing the basic iterative method (BIM)
~\cite{kurakin2018adversarial}, 
also called I-FGSM, which iteratively  applies FGSM 
with a small step size.
Madry et al. refined  BIM and introduced the projected 
gradient descent (PGD) method~\cite{madry2017towards}.
Unlike BIM, which starts from an original input, 
PGD initializes at a random point 
and 
continuously performs
random attacks.
Despite this distinction,
these two methods are often considered identical.

Moosavi-Dezfooli et al. proposed the DeepFool
method~\cite{moosavi2016deepfool}, which
seeks the smallest amount of perturbation
for a successful
attack.
Unlike FGSM and similar methods 
that require the manual perturbation 
parameter setting,
DeepFool treats the decision boundary as linear when the perturbation
distance is minimal.
Under this scenario, the orthogonal vector is derived by linearizing the
decision boundary using the Taylor expansion and seeking AEs along the
orthogonal vector.

Carlini et al. introduced an attack method known as CW, which frames
AE generation as an optimization problem aimed at minimizing
the difference between 
unattacked inputs and AEs~\cite{carlini2017towards}.
CW achieves minimal perturbations and demonstrates a high attack 
success rate.

Papernot et al. proposed maximal Jacobian-based saliency map 
attack (JSMA)~\cite{papernot2016limitations}.
This method calculates a Jacobian matrix, and based on this, derives an 
adversarial saliency map.
A greedy algorithm subsequently selects the pixel with the highest
value in the adversarial saliency map, which is then perturbed.
These steps are iterated until the maximum number of perturbed pixels
is reached, ultimately generating an AE.

\subsubsection{Score-based attack}

Narodytska et al. introduced a score-based attack method named
LocalSearch (LS)~\cite{narodytska2016simple}, which
generates AEs by minimizing 
the prediction probability of the original label.
Through a greedy local search, it generates local neighborhood images
perturbed by a few pixels from the original input.
Subsequently, it selects the image with the lowest predicted
probability of the original label.
These steps are iterated until the predicted label of the perturbed image
is changed from the label of the original one.

\subsubsection{Decision-based attack}
Brendel et al. introduced a decision-based attack (DBA) named Boundary
Attack, which operates under the assumption that only predictive
labels are provided.
This method minimizes the perturbation amount by approaching the
original input along the decision boundary while maintaining image
misclassification~\cite{brendel2017decision}.

Chen et al. proposed the HopSkipJumpAttack (HSJA) method, which estimates the 
decision boundary gradient by approximating the local 
decision boundary using a Monte Carlo method
~\cite{chen2020hopskipjumpattack}.
HSJA estimates the direction orthogonal to the decision 
boundary surface from the region near the AE and minimizes the 
perturbation amount
in combination 
with a binary search.

\subsection{Adversarial defense}

Real-world systems employing DNN models encounter the risk of adversarial
attacks from malicious entities incentivized to cause harm.
Consequently, various adversarial defense methods have 
been developed
to
protect
systems from such attacks.
Given the unpredictable randomness inherent in many real-world
environments, assessing the robustness of a system against AEs is a
test for its resilience under worst-case scenarios.

Adversarial defense methods for systems utilizing DNNs are typically
categorized into three:
adversarial training~\cite{goodfellow2014explaining, madry2017towards,
  shafahi2019adversarial, zhang2019theoretically}, 
input transformation~\cite{xie2017mitigating, dziugaite2016study,
  meng2020athena, guo2017countering, buckman2018thermometer}, 
and detection methods~\cite{feinman2017detecting,
  ma2018characterizing, wang2020dissector, wang2019adversarial,
  zhao2021attack, rottmann2020detection, chen2021attackdist}.

Among them, adversarial training is the most prevalent approach, 
which strengthens the 
systems' resilience against AEs by incorporating them into the
training data.
However, 
although effective, 
these methods often come at
the expense of reduced accuracy for benign samples and increased
computational overhead.

Input transformation offers an avenue to mitigate the impact of AEs by
preprocessing the input data.
In tasks such as image classification, R\&P
transform~\cite{xie2017mitigating} and JPEG
transform~\cite{dziugaite2016study} are employed to alter input
images.
Nevertheless,
the uniform application of transformations across all
inputs may distort benign samples, thereby diminishing classification
accuracy.
Moreover, preprocessing methods must be tailored to the input 
data types of DNNs including images, audio, and text.

In contrast to other two approaches,
detection methods accurately identify benign samples while posing
challenges in tasks requiring the precise categorization of the
original input such as sign recognition in autonomous
driving~\cite{cirecsan2012multi}.

\subsection{Adversarial defense using AE vulnerabilities}

Attack as defense ($\rm A^2D$) is a defense method that targets AE
vulnerabilities.
If AEs are near the decision boundary in the feature space and are
subjected to another attack, they can easily traverse it, altering the
classification result~\cite{zhao2021attack}.
AE vulnerability is measured by re-attacking the input data that may
be adversarial, employing an iterative search attack, and assessing
the re-attack costs, i.e., costs associated with the number of
iterations needed to alter the identification result.
AEs can be identified based on the disparity in the attack costs
required to change their category between AEs and benign samples in
training samples (prepared separately based on cases to be attacked).
BIM~\cite{kurakin2018adversarial},
JSMA~\cite{papernot2016limitations}, and
DBA~\cite{brendel2017decision} are utilized for iterative attacks,
while the k-nearest neighbor (k-NN) algorithm or standard score
(Z-score) is employed to differentiate AEs based on the attack costs.

The Attackdist method is a detection approach that operates on two primary
assumptions.  
First, adversarial perturbations generated by the attack algorithms
must be close to the optimal solution.
Second, the optimal solution is near the decision
boundary~\cite{chen2021attackdist}.
If an AE is re-attacked, the perturbation should be substantially
smaller than that in benign samples.
This method utilizes the Lp norm of the adversarial perturbation for
detection.

Another detection method leverages the CW attack method based on two
fundamental principles.
First, perturbation caused by a CW attack is minimized through iteration, 
thereby bringing it close to the decision boundary.
Second, perturbation from a CW re-attack is much smaller for the
already attacked image than that for the original
image~\cite{rottmann2020detection}.
This method discriminates AEs generated through CW attacks, which are
challenging to detect, by conducting additional CW re-attacks, with
discrimination based on the number of iterations.

Despite the high discrimination ability exhibited by the
aforementioned methods for AE detection through re-attacking input
images, they focus solely on AE detection and do not consider the
identification of the correct original input class.

\subsection{Recent adversarial defense methods}
\label{ssec:recent_methods}

Among the state-of-the-art research on adversarial defense, in this
section, we present some recently emerging ideas similar to ours.

Salman et al. introduced the unadversarial method, which employs an
inverse gradient for AE ~\cite{salman2021unadversarial}.
Their approach improves a model's performance and robustness to
corrupted images by generating unadversarial examples (un-AEs) that
minimize losses instead of adversarial perturbations.

More recently, Chen et al. proposed the adversarial visual prompting
(AVP) method, which enhances adversarial robustness through visual
prompting~\cite{chen2023visual}.
AVP improves robustness during testing by designing prompting to
correct AE classification results in advance.

Wang et al. introduced the FeConDefense
method~\cite{wang2023fecondefense}, building upon previous study on
 a reverse attack method~\cite{mao2021adversarial} by Mao et al.
These two techniques utilize pseudo loss gradients with contrastive
loss and feature consistency loss to incorporate reverse perturbations
into AEs, thereby restoring natural images.

The concept of our proposed method aligns with those of the
aforementioned approaches, which utilize gradients for AEs to
introduce perturbations in the inverse direction of adversarial
perturbations.
However, while state-of-the-art methods improve model robustness, they
differ substantially from the objectives of our study.
Un-AEs, for instance, address domain shifts to enhance the robustness against
corrupted images without directly improving adversarial robustness.
Meanwhile, AVP applies identical prompting to benign samples and AEs,
severely decreasing the classification accuracy of the former.
Similarly, reverse attack leads to a notable decrease in the
classification accuracy for benign samples, while details regarding
the classification accuracy obtained with FeConDefense are limited.
Furthermore, 
AVP and FeConDefense are only effective against gradient-based
attacks, which are necessary for training the defense model, and they
do not assess defense performance against various adversarial attacks.
Additionally, these methods are categorized as input transformation
methods, while our approach serves as a postprocessing technique for
the detector, thereby avoiding deterioration 
in
the classification
accuracy for benign samples.

Recently, the fields of natural language and speech processing have been
extensively researched with respect to AE detection.
Methods such as
frequency-guided word substitutions
(FGWS)
~\cite{mozes2020frequency},
TextFirewall~\cite{wang2021textfirewall},
word-level differential reaction%
~\cite{mosca2022suspicious}, and 
adversary detection with data and model uncertainty%
~\cite{yin2022addmu} in natural language processing (NLP), as well as
acoustic-decoy~\cite{KWON2020357} and 
FraudWhisler~\cite{wang2024fraudwhistler}
in speech processing,
focus
solely on detecting AEs.
However, contemporary approaches include mechanisms for
correcting the detected AE labels.
For instance, 
randomized substitution and vote
(RS\&V)
~\cite{wang2022detecting} generates multiple
similar sentences by substituting synonyms in AEs and detects them
based on the consistency of their classification results, which also
enables the correction of their labels.
Reactive perturbation defocusing
 (RAPID)
~\cite{yang2023best} proposes a method that combines a detector with a
perturbation focusing on a rectifier and uses pseudo-semantic
filtering as a post-process to identify and correct the AE labels.
This method requires training a neural network for the detector and
utilizing the rectifier in response to it.
Nevertheless, these approaches are limited to natural language modalities and 
depend on specific detectors for functionality, distinguishing them
from our method that is adaptable to any detector.

\subsection{Rectification of AEs}

Few studies have explored rectifiers that integrate with any detector,
such as the proposed method,
with the method proposed by Kao et al.
being a rare example~\cite{kao2022rectifying}.
They introduced a rectification method that addresses the limitations
of the existing defense approaches.
Their research was the first to focus on the requirements of various
postprocessing techniques for revealing the correct classification
result of AEs detected by the detector.
They explored the feasibility of rectifying AEs and restoring them to
their correct states by modifying or removing the identified regions
of interest, estimated using XAI.
This method 
does not require input exclusion or 
computationally intensive processes, thereby mitigating the limitations
associated with the current defense methods.
Moreover, this method outperforms the four baselines (Autoencoder for
denoising, JPEG compression, full-image Gaussian blur, and random
pixel replacement in an image) that are
implemented as methods to rectify detected AEs.
However, its success rate for rectification  substantially
depends on the attack method used for AE creation.
Additionally, a limitation of this method is that it can only rectify
AEs against DNNs that can utilize XAI methods.

Some of the latest methods for NLP
DNNs described in Section \ref{ssec:recent_methods} primarily focus on
detecting AEs, yet they also provide AE rectification.
RS\&V~\cite{wang2022detecting}, for instance, applies perturbations through synonym
substitution to both detect and rectify AEs.
It generates multiple distinct perturbation patterns for an input,
feeds these perturbed inputs into a DNN model, and then takes a majority
vote of the resulting labels.
If the majority label matches the original input's label, the input is
deemed benign; if the majority label changes, the input is identified
as an AE, and the majority label becomes the rectified result.

RAPID is a defense approach consisting of a learning-based detector
and a semantic rectifier based on perturbation defocusing.
The rectifier applies safe perturbations to a detected AE to
neutralize adversarial perturbations and generate sentences that
retain meaning close to the original input.
Specifically, synonym replacement is performed based on word
significance and classification probability by word replacement,
ensuring the semantics remain unchanged.

FGWS~\cite{mozes2020frequency} employs synonym replacement to detect
AEs and to predict correct labels similar to RAPID.
This method applies perturbations based on word frequency
characteristics.

The above methods 
are specifically designed for DNNs
in NLP.
Furthermore, methods such as RAPID and FGWS require domain-specific data
and model training, unlike our proposed method, which requires no such
preparation and can easily combine with various detectors.

\begin{figure*}[t]
  \centering
  \begin{tabular}{@{}p{5.5cm}@{~~~}p{5.5cm}@{~~~}p{5.5cm}@{}}
  \includegraphics[width=5.5cm]{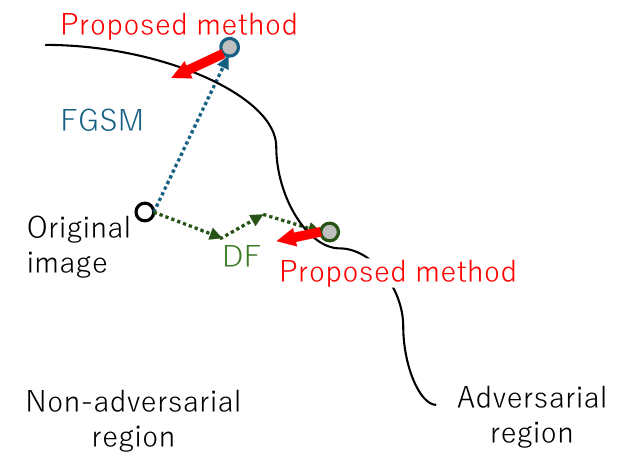} &
  \includegraphics[width=5.5cm]{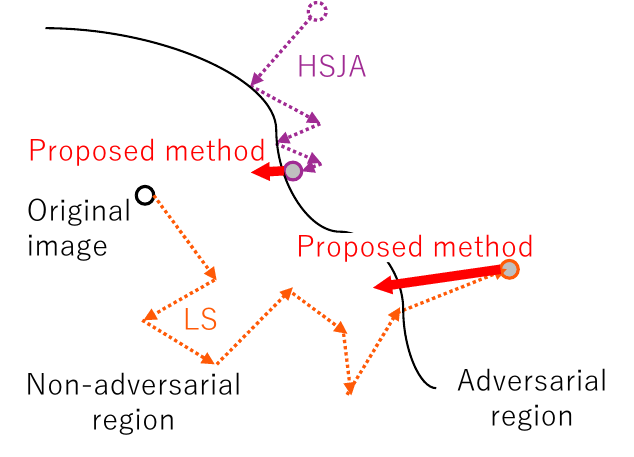} &
  \includegraphics[width=5.5cm]{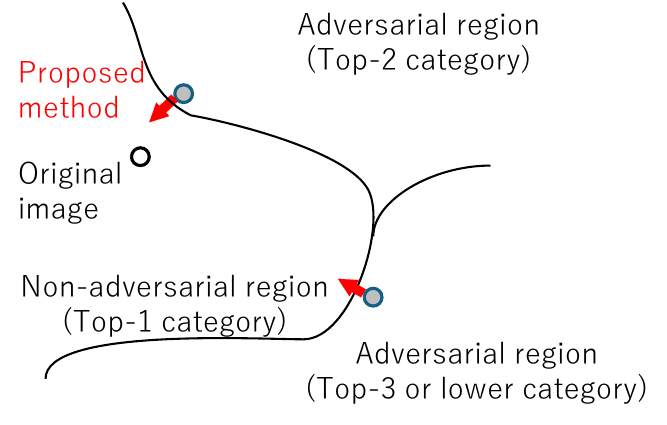} \\
  {\footnotesize (a) Rectifying AEs generated by white-box attacks, such as FGSM~\cite{goodfellow2014explaining} and DF~\cite{moosavi2016deepfool}.} &
  {\footnotesize (b) Rectifying AEs generated by black-box attacks, such as LS~\cite{narodytska2016simple}  and HSJA~\cite{chen2020hopskipjumpattack}.} &
  {\footnotesize (c) Rectifying AEs generated by targeted attacks.} \\
  \end{tabular}
  \caption{Conceptual 
interpretation
 of the proposed method.}
  \label{fig:idea}
\end{figure*}

\section{Proposed AE rectification method}
\label{sec:Methodology}
\begin{figure}[t]
  \centering
  \includegraphics[width=85mm]{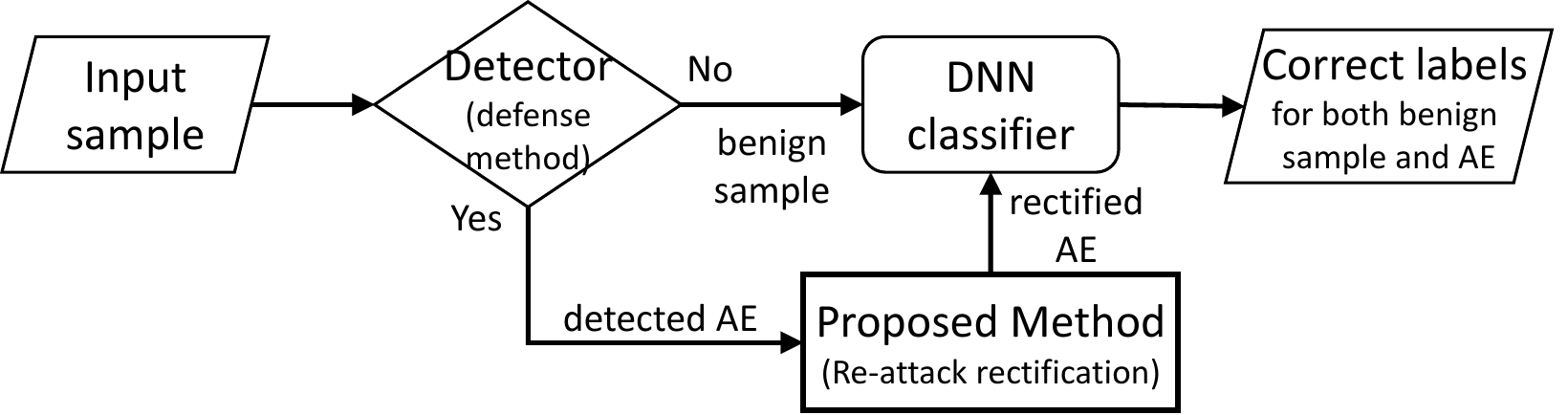}
  \caption{Relationship between our proposed method and AE detection method.}
  \label{fig:flow}
\end{figure}

\subsection{Key idea}

The essence of the proposed  AE rectification method is the estimation of
the correct label of benign samples by re-attacking the AE
identified by the defense method.
Because white-box attack methods create AEs according to the gradients
of the loss function, AEs are located near decision boundaries in the
adversarial regions.
Importantly, 
note that unlike standard adversarial attacks that create AEs,
re-attacks are conducted without any knowledge regarding the correct
category or original input.
By applying perturbations in the opposite direction to 
AE generation, 
reverting these examples
back to non-adversarial regions across the boundary is possible. 
Thus, our rectification method leverages the gradient of the loss
function derived from the target defense model to shift AEs toward
decreasing the confidence of the misidentified category, thereby
predicting their correct labels.

The proposed method is designed specifically to correct AEs on its own,
enabling its integration with any arbitrary AE detector.
It operates under the assumption that it will only receive AEs
identified by the detector as inputs, excluding benign samples.
Thus,
the proposed method can continuously re-attack until the AE label
changes.
This design eliminates the need for task-specific preliminary trainings
or parameter adjustments, 
which is a notable advantage of the proposed method.

Utilizing re-attacks for rectifying AEs, the proposed method offers
broad applicability across various tasks and data modalities.
Although
alternatives such as smoothing or noise addition/removal can
rectify AEs, specific preprocessing methods tailored to the input data type
for DNNs, including images, sound, and video, remain necessary.
However, our method does not require such preliminary processes or
adjustments, and is independent of the input data type for DNN.
Moreover, it
applies to any DNNs where adversarial attack methods exist.

\subsection{Conceptual interpretation
  in the feature space}
\label{ssec:concept}

Figure~\ref{fig:idea} illustrates
the functioning of the  proposed method against various attack
strategies.
AEs created by FGSM involve a single calculation of the gradient of
the input, resulting in larger perturbations (Figure~\ref{fig:idea}(a)),
in contrast to methods such as BIM and DeepFool~\cite{moosavi2016deepfool}
that produce AEs with smaller perturbations through iterative processes.
Our method, utilizing a white-box attack method for re-attacking, moves
AE toward a direction that reduces the confidence of the
misclassified category, transforming it into a non-adversarial sample.
Re-attack using the same white-box attack method as that used to
create AE, can possibly yield successful label correction.
Furthermore, even if methods used for re-attacking differ from the
original attack strategies employed for AE creation through white-box
attack methods, the proposed method remains effective in terms of
employing gradients to shift AEs toward a reduction in the confidence
of incorrectly recognized categories, enabling the restoration of
their correct categories.

Score-based black-box attacks that do not utilize
the
loss function gradient of the target defense model, such as LS,
generate AEs at locations relatively far from the decision boundary,
making their rectification challenging, as shown in
Figure~\ref{fig:idea}(b).
However, because our method assumes that all inputs are AEs, it allows for
re-attacking until the classification result of the AE changes.
This enables the method to continuously re-attack AEs created by LS
until they cross the decision boundary, even potentially correcting
AEs that are far from the boundaries back to their original categories.

Decision-based black-box attacks, such as HSJA, typically start
searching near the adversarial region and include a binary search
toward the original input, as shown in Figure~\ref{fig:idea}(b).
As a result, AEs generated by HSJA end up close to the boundary that
separates the adversarial and non-adversarial regions, similar to AEs
generated by white-box attacks.
Hence, even if AEs are generated without using the loss gradient, our
re-attack method can still effectively correct their labels.

Conversely,
rectifying AEs created by targeted attacks that misclassify them into
significantly less confident categories becomes increasingly difficult
owing to large perturbations.
Nevertheless, our method is expected to correct these AEs effectively
because it continues re-attack until the AE classification result
changes.
The key to successful correction, especially for AEs aimed to be
recognized within the least confident Top-3 categories or lower,
depends on the proximity between the misclassified and correct
(Top-1) category areas.
As illustrated in Figure~\ref{fig:idea}(c), successful label correction
can be realized using
our method if there is an adequate boundary region
that connects
the misclassified and correct categories.
Given the high dimensionality of DNN inputs that encompass numerous
categories, the proposed approach is believed to be effective against
targeted attacks as numerous categories are expected to be adjacent to
non-adversarial regions.

\subsection{Theoretical foundations}
\label{ssec:theoretical_analysis}
This section discusses the theoretical perspectives through which the
proposed method is capable of rectifying AEs using a white-box attack
method.
Adversarial attack methods typically create a minimal adversarial
perturbation ${\bm \delta}$ that changes the classification result by
solving the following optimization problem:
\begin{displaymath}
  \text{minimize}~ \|{\bm \delta}\|_p, \;\;\; ~ 
  \; s.t.\;\; C({\bm x}+{\bm \delta}) \neq C({\bm x})
\end{displaymath}
The optimized perturbation ${\bm \delta^*}$, which is  minimized
while residing in an adversarial region,
is very close to the decision boundary of the original
classification region (non-adversarial region).
Therefore, if the 
AE~ ${\bm x} + {\bm \delta^*}$ moves even
slightly toward the original input, it will be classified as the
original class, i.e.,
\begin{displaymath}
  C\left({\bm x} + (1- \mu){\bm \delta^*}\right) 
  = C({\bm x}), \;\;\; 
  ~ \mu \gtrsim 0
\end{displaymath}
where 
$\mu \gtrsim 0$
indicates that $\mu$ is approximately equal to 0 but greater than 0.

Here, we consider re-attacking the AE that incorporates an ideal
perturbation ${\bm \delta}^*$ as described above.
Specifically,
\begin{displaymath}
  \text{minimize}~ \|{\bm \delta'}\|_p, \;\;\; ~ \; s.t.\;\; C({\bm x}+{\bm \delta^*}+{\bm \delta'}) \neq C({\bm x}+{\bm \delta^*})
\end{displaymath}
The approximate solution ${\bm \delta'^*}$ of the above problem
can be regarded as $-\mu {\bm \delta^*}$.
\begin{displaymath}
  \| -\mu {\bm \delta^*}\|_p \approx 0, \;\;\; ~ C({\bm x}+{\bm \delta^*}-\mu {\bm \delta^*}) = C({\bm x}) \neq C({\bm x}+{\bm \delta^*})
\end{displaymath}
Therefore, rectifying an AE requires determining the direction $-{\bm
  \delta^*}$ and the magnitude $\mu$, under the condition that
the original input ${\bm x}$
remains unknown.

Here, we assume that ${\bm x}+{\bm \delta^*}$ is sufficiently close to
the decision boundary of the original class $C({\bm x})$
and that the logits (i.e., pre-softmax outputs) for classes other than
$C({\bm x})$ and $C({\bm x}+{\bm \delta^*})$ are sufficiently low.
These assumptions enable us to regard the classification
as a binary classification
problem.
Furthermore, assuming that the entire model can be regarded as linear
and employs a binary cross entropy loss with no regularization term,
the direction of ${-\bm \delta^*}$ can be obtained by calculating the
direction that decreases the logit of $C({\bm x}+{\bm \delta^*})$ at
${\bm x}+{\bm \delta^*}$.
\begin{displaymath}
  - {\bm \delta^*} = \lambda 
                    \nabla L({\bm \theta},{{\bm x}+{\bm \delta^*}},C({\bm x}+{\bm \delta^*})), 
  ~ \lambda > 0
\end{displaymath}
where $L({\bm \theta}, {\bm x} + {\bm \delta^*}, C( {\bm x} + {\bm
  \delta^*}))$ denotes a loss function of the binary classifier with
parameter ${\bm \theta}$ for an input ${\bm x + {\bm \delta^*}}$ and
its target $C({\bm x} + {\bm \delta^*})$.
This means that it is possible to rectify the AE and estimate the
class $C({\bm x})$ of the original input ${\bm x}$ by re-attacking the
AE with a white-box attack such as FGSM or BIM.

Note that precise estimation for the direction of $-{\bm \delta}^*$
is not necessary as long as it directs towards the region of $C({\bm
  x})$.
However, factors such as 
a large $\mu$, 
significant logits from other classes than $C({\bm x})$ and $C({\bm x} + {\bm \delta}^*)$,
 and complex decision boundaries 
increase deviations from these assumptions.
The further an AE 
deviates from these assumptions in the feature space,
the more challenging it becomes to accurately
estimate the direction
of $-{\bm \delta}^*$, thereby making the rectification of the AE more
difficult.

\subsection{Process flow}
\label{ssec:process_flow}
The interplay between the proposed AE rectification method and
conventional defense methods is illustrated in Figure~\ref{fig:flow}.
Our proposed method involves re-attacking an AE identified via existing
detection methods to deduce the correct class of its original
image.
Without limitations, it is compatible with various adversarial attack
methods encompassing white-box and black-box attacks.
Nevertheless, this study focuses on white-box attacks,
considering that defenders frequently possess permissions to access
the internal information of the defended DNNs.
While the proposed method requires no prior adjustment of the
perturbation amount, computing the perturbation direction based on an
input sample and a target defense model is crucial.
Moreover, utilizing a white-box attack that calculates the loss
function gradient for re-attacks allows for the most effective optimal
direction estimation,
as discussed in Section~\ref{ssec:theoretical_analysis}.

\subsection{Re-attack methods}
\label{ssec:reattack_methods}

Although the proposed method can
utilize any attack method for re-attacking AEs,
given the availability of the internal information regarding the DNN model,
we opt to employ FGSM~\cite{goodfellow2014explaining},
BIM~\cite{kurakin2018adversarial}, and
DeepFool~\cite{moosavi2016deepfool} methods for AE re-attacks in this paper.
The proposed method can employ any white-box attack method.
To demonstrate that even simple methods suffice for effective
rectification, we selected FGSM, BIM, and DF, the simplest and most
distinct white-box algorithms.

\subsubsection{Re-attack with FGSM}
\label{ssec:fgsm}
FGSM re-attacks the detected AE ${\bm x_{a}}$ by adding 
one-step perturbation in the gradient direction.
However, unlike an original FGSM, which does not iterate processes, we
incorporate a linear search to FGSM to determine the amount of
movement, thereby ensuring that it can detect samples whose labels
change.
Notably, FGSM calculates the gradient only once, 
differing from iterative optimization methods such as 
BIM and PGD, even after employing the linear search.
\begin{algorithm}[t]
  \caption{Re-attack with FGSM}
  {
  \label{alg:fgsm}
  \begin{algorithmic}[1]
      \REQUIRE detected AE ${\bm x_{a}}$, perturbation size $\epsilon$, iterations $s$
      \ENSURE re-attacked AE ${\bm x'_{a}}$
      \STATE  ${\epsilon_{s}}\leftarrow\epsilon/s$
      \STATE  $g \leftarrow\nabla_{\bm x_{a}} L(\theta,{\bm x_{a}},y_{a})$
      \STATE  $\epsilon\leftarrow0$, $i\leftarrow0$
      \WHILE{$C(\bm x'_{a})\neq y_{a} ~~\OR~~ i < s$}
        \STATE  $\epsilon\leftarrow\epsilon+{\epsilon_{s}}$
        \STATE  ${\bm x'_{a}}\leftarrow{\bm x_{a}}+\epsilon \cdot \rm{sign}(g)$
        \STATE  $i\leftarrow i+1$
        \ENDWHILE
      \RETURN ${\bm x'_{a}}$
  \end{algorithmic} 
  }
\end{algorithm}
Algorithm~\ref{alg:fgsm} outlines the detailed algorithm, where
perturbations are calculated as follows:
\begin{displaymath}
{\bm x'_{a}}={\bm x_{a}}+\epsilon \cdot \rm{sign}(\nabla_{\bm x_{a}} L(\theta,{\bm x_{a}},y_{a}))
\end{displaymath}
where ${\bm x'_{a}}$ represents the image after re-attack, 
$y_{a}$ denotes the $\bm x_{a}$ label, $\epsilon$ is the 
parameter controlling the perturbation size, $\theta$ signifies the 
model parameter, $L$ denotes the 
loss function, and $\rm{sign}$ is the sign 
function.

\subsubsection{Re-attack with BIM}
BIM re-attacks the detected AE ${\bm x_{a}}$ by adding several 
perturbations in the gradient direction with a small step size.
\begin{algorithm}[t]
  \caption{Re-attack with BIM}
  {
  \label{alg:bim}
  \begin{algorithmic}[1]
      \REQUIRE detected AE ${\bm x_{a}}$, perturbation size $\epsilon$, step size $\alpha$, iterations $N$
      \ENSURE re-attacked AE ${\bm x'_{a}}$
      \STATE ${\bm x'_{a}}_{(0)}={\bm x_{a}}$
      \FOR{$n=0,\ldots,N-1$}
          \STATE  ${\bm x'_{a}}_{(n+1)}\leftarrow{\bm x'_{a}}_{(n)}+\alpha \cdot \rm{sign}(\nabla_{\bm x'_{a}} L(\theta,{\bm x'_{a}}_{(n)},y_{a}))$
          \STATE  ${\bm x'_{a}}_{(n+1)}\leftarrow \rm{Clip}_{{\bm x_{a}},\epsilon}({\bm x'_{a}}_{(n+1)})$
      \ENDFOR
      \RETURN ${\bm x'_{a}}_{(N)}$
  \end{algorithmic} 
  }
\end{algorithm}
Algorithm~\ref{alg:bim} outlines the detailed algorithm, where
perturbations are calculated as
\begin{eqnarray}
{\bm x'_{a}}_{(0)}&=&{\bm x_{a}},  \\
{\bm x'_{a}}_{(n+1)}&=& \mbox{Clip}_{{\bm x_{a}},\epsilon}({\bm x'_{a}}_{(n)}\\ \nonumber
                   & &+\alpha \cdot \rm{sign}(\nabla_{\bm x_{a}} L(\theta,{\bm x'_{a}}_{(n)},y_{a})))
\label{eq:bim}
\end{eqnarray}
where
$\alpha$ is the step 
size. 
After updating ${\bm x}'_{{adv}_{(n)}}$,
$\rm{Clip}$ function is applied to clip AEs, optimizing them within the
$\epsilon$ region of the original input.

\subsubsection{Re-attack with DeepFool}
DeepFool re-attacks a detected AE ${\bm x}_{a}$
by estimating decision boundaries for all
classes from an original input.
It calculates a perturbation toward
classes with the nearest boundary to the original.
\begin{algorithm}[t]
  \caption{Re-attack with DeepFool}
  {
  \label{alg:df}
  \begin{algorithmic}[1]
      \REQUIRE detected AE ${\bm x_{a}}$, steps $N$
      \ENSURE re-attacked AE ${\bm x'_{a}}$
      \STATE ${\bm x'_{a}}_{(0)}\leftarrow{\bm x_{a}}$
      \STATE $i\leftarrow0$
      \WHILE{$\hat{k}({\bm x'_{a}}_{(i)})= {\hat{k}(\bm x_{a})}  ~~\OR~~ i < N$}
          \FOR{$ k:k\neq\hat{k}({\bm x_{a}}) $}
              \STATE  $ {{\bm w}'_{k}}\leftarrow \nabla f_{k}({\bm x'_{a}}_{(i)}) - \nabla f_{\hat{k}({\bm x_{a}})}({\bm x'_{a}}_{(i)})$
              \STATE  ${f'_{k}}\leftarrow f_{k}({\bm x'_{a}}_{(i)}) - f_{\hat{k}({\bm x_{a}})}({\bm x'_{a}}_{(i)})$
          \ENDFOR
          \STATE $\hat{l}\leftarrow\argmin_{  k:k\neq\hat{k}({\bm x_{a}}) }\frac{|{f'_{k}}|}{{\|{{\bm w}'_{k}}\|}_{2}}$
          \STATE ${\bm r_{i}}\leftarrow\frac{|{f'_{\hat{l}}}|}{{\|{
          {\bm w}'_{\hat{l}}}\|}^{2}_{2}}{{\bm w}'_{\hat{l}}}$
          \STATE ${\bm x'_{a}}_{(i+1)}\leftarrow{\bm x'_{a}}_{(i)}+{\bm r_{i}}$
          \STATE $i\leftarrow i+1$
      \ENDWHILE
      \RETURN ${\bm x'_{a}}_{(N)}$
  \end{algorithmic} 
  }
\end{algorithm}
The detailed algorithm is shown in Algorithm~\ref{alg:df}.
To handle non-linear boundaries, the DeepFool method iterates the
linear approximation of  boundaries and 
the addition of
the smallest perturbation 
to the nearest class.

The following describes the re-attack algorithm using DeepFool.
Let $f(\cdot)$ be a classifier,
and define 
the classification $\hat{k}({\bm x})$ as:
\begin{equation}
\hat{k}({\bm x})=
\mathop{\rm{arg~max}}\limits_k~
f_{k}({\bm x})
\label{eq:df_1}
\end{equation}
where $f_{k}({\bm x})$ is the output score of $f({\bm x})$ corresponding to class $k$.

At each iteration, 
for the current input ${{\bm x}'_{a}}_{(i)}$,
DeepFool calculates the output score $f_k ({{\bm x}'_{a}}_{(i)})$ and
its corresponding gradient $\nabla f_k ({{\bm x}'_{a}}_{(i)})$ for each class $k$.
Subsequently, for every other class $k \neq \hat{k}({\bm x}_{a})$, it approximates
the distance from ${{\bm x}'_{a}}_{(i)}$ to the decision boundary between class $\hat{k}({\bm x}_{a})$ and
class $k$ by computing 
$  |f'_k| / \| {\bm w}'_k \| $,
where ${\bm w}'_k$ and $f'_k$ are calculated as follows:
\begin{eqnarray} 
{\bm w}'_{k} &=& \nabla f_{k}({{\bm x}'_{a}}_{(i)}) - \nabla f_{\hat{k}({\bm x}_{a})}({{\bm x}'_{a}}_{(i)}) \\
f'_{k}       &=& f_{k}({{\bm x}'_{a}}_{(i)}) - f_{\hat{k}({\bm x}_{a})}({{\bm x}'_{a}}_{(i)}) 
\label{eq:df_2}
\end{eqnarray}
Next, it identifies the class $\hat{l}$ with the nearest decision
boundary to ${{\bm x}'_{a}}_{(i)}$, i.e.,
\begin{equation}
\hat{l} = 
\mathop{\rm{arg~min}}\limits_{k:k\neq\hat{k}({\bm x}_{a})}~
\frac{|{f'_{k}}|}{{\|{{\bm w}'_{k}}\|}_{2}}
\end{equation}
It then computes a vector that projects ${{\bm x}'_{a}}_{(i)}$ onto the
hyperplane approximating the decision boundary between of class
$\hat{l}$ and $\hat{k}({{\bm x}'_{a}}_{(i)})$,
yielding 
the minimal perturbation ${\bm r_{i}({{\bm x}'_{a}}_{(i)})}$ calculated as follows:
\begin{equation}
{\bm r}_{i}
= \frac{|f'_{\hat{l}}|}{\|{\bm w}'_{\hat{l}}\|^{2}_{2}} {\bm w}'_{\hat{l}}
\label{eq:df_3}
\end{equation}
By adding ${\bm r}_i$ to ${{\bm x}'_{a}}_{(i)}$, DeepFool updates
the input and obtains ${{\bm x}'_{a}}_{(i + 1)}$.

DeepFool repeats this process 
with incrementing $i$ 
until reaching the iteration limit or until ${{\bm x}'_{a}}_{(i)}$ lies
outside the region of $\hat{k}({\bm x_{a}})$ .

\section{Evaluation}
\label{sec:Evaluation}
To assess the efficacy of the proposed method,
four
experimental tests were executed as outlined below.
Initially, the effectiveness of the proposed method with image
classification DNN models against various attack methods was validated
(Experiment 1).
Experiment 1 was conducted under an untargeted attack scenario against
white-box and black-box attack methods (Experiment 1a) and a targeted
scenario (Experiment 1b).
Subsequently, 
comparative analyses with the
state-of-the-art rectification methods reported in
Refs.~\cite{kao2022rectifying,wang2022detecting} 
were performed (Experiments 2a and 2b).
Consequently, the defense performance of the proposed method was
further illustrated in conjunction with the detector outlined in the
$\rm A^2D$ method~\cite{zhao2021attack} (Experiment 3).
Finally, to demonstrate the applicability to other data modalities, 
the proposed method was applied to speech recognition (Experiment 4).

\begin{table}[t]
  \centering
  \caption{Re-attack parameters for rectification in the proposed method.}
  \label{tbl:proposed_method_parameter}
  {
  \begin{tabular}{c|l}
  \hline
  \multicolumn{1}{c|}{Re-attack method} & \multicolumn{1}{c}{Parameter} \\
  \hline
  \multicolumn{1}{c|}{FGSM} & \multicolumn{1}{c}{
    $s=1,000$,
    $\epsilon=1.0$
  } \\
  \hline
  \multicolumn{1}{c|}{\multirow{1}{*}{BIM}} & \multicolumn{1}{c}{\multirow{1}{*}{
    $\epsilon=0.3$, 
    $\alpha=0.05$, 
    $N=10$
  }} \\
  \hline
  \multicolumn{1}{c|}{\multirow{1}{*}{DF}} & \multicolumn{1}{c}{\multirow{1}{*}{
    $s=100$
  }} \\ 
  \hline
  \end{tabular}
  }
\end{table}

\subsection{Experiment 1a: Rectification performance against various attack methods including black-box attacks (untargeted attack)}
\label{ssec:untargeted}

\subsubsection{Setup}

In this experiment, we assessed the rectification performance of our
method by combining various datasets and attack methods under an
untargeted attack scenario.
First, we applied it to AEs 
generated by white-box attacks including 
FGSM~\cite{goodfellow2014explaining}, BIM~\cite{kurakin2018adversarial}, 
DeepFool (DF)~\cite{moosavi2016deepfool}, CW~\cite{carlini2017towards}, 
and JSMA~\cite{papernot2016limitations}.
Subsequently,
we utilized AEs generated by
black-box attacks such as LocalSearch (LS)~\cite{narodytska2016simple} 
and HopSkipJumpAttack (HSJA)~\cite{chen2020hopskipjumpattack}.
These attack methods were chosen based on the guidelines for defense
evaluation~\cite{carlini2019evaluating}, ensuring diversity and
representation while avoiding the use of similar methods.
We employed the implementations of the attack methods in the FoolBox
framework~\cite{rauber2017foolbox}.
Attack parameters for creating AEs using the seven methods were
configured as the default values of FoolBox.
For re-attacking, we selected FGSM, BIM, and DF, based on the reasons
given in Section~\ref{ssec:reattack_methods}.
The re-attack parameters of the proposed method were
configured as the default values of FoolBox, and
are
 listed in
Table \ref{tbl:proposed_method_parameter}.

For Experiments 1 and 3, we employed three image datasets: MNIST
~\cite{deng2012mnist}, CIFAR-10~\cite{krizhevsky2009learning}, and
ImageNet-1000 (ILSVRC2012)~\cite{imagenet}.
Classification models for MNIST and CIFAR-10 were implemented 
based on previous studies~\cite{kao2022rectifying} to fairly compare our method
with the previous one used in Experiment 2, while
VGG-19~\cite{simonyan2014very} served as the ImageNet 
classifier.

For each combination of the three datasets and seven attack methods,
1,000 samples were selected 
for which the classification model 
correctly identified the original input, and the adversarial 
attack succeeded.
We defined 
the percentage of the
rectification success rate as an evaluation criterion, 
using which
the result of identifying AEs after 
rectification matched that of the original input.

\begin{table}[t]
  \centering
  \caption{Experiment 1a: Success rates of rectification (untargeted attack).}
  \label{tbl:attack_comparison}
  {
  \begin{tabular}{@{~}c@{~}||@{~}c@{~}|@{~}c@{~}c@{~}c@{~}c@{~}c@{~}|@{~}c@{~}c@{~}}
  \hline
  \multirow{3}{*}{Dataset} &  
  \multirow{2}{*}{Re-attack} & 
  \multicolumn{7}{@{~}c@{~}}{\multirow{1}{*}{Attack method}} \\
  \cline{3-9}
  &
  \multirow{2}{*}{method}& 
  \multicolumn{5}{@{~}c@{~}|@{~}}{\multirow{1}{*}{White-box}} &
  \multicolumn{2}{@{~}c@{~}}{\multirow{1}{*}{Black-box}} \\
  \cline{3-9} 
  & \multicolumn{1}{@{~}c@{~}|@{~}}{}
  & \multicolumn{1}{@{~}c@{~}}{FGSM} 
  & \multicolumn{1}{@{~}c@{~}}{BIM} 
  & \multicolumn{1}{@{~}c@{~}}{DF} 
  & \multicolumn{1}{@{~}c@{~}}{CW} 
  & \multicolumn{1}{@{~}c@{~}|@{~}}{JSMA} 
  & \multicolumn{1}{@{~}c@{~}}{LS} 
  & \multicolumn{1}{@{~}c@{~}}{HSJA} \\
  \hline\hline
  \multirow{3}{*}{MNIST} 
  &\multicolumn{1}{@{~}c@{~}|@{~}}{FGSM} & \textbf{0.999} & \textbf{0.999} & \textbf{0.978} & \textbf{1.000} & 0.993 & 0.938 & \textbf{1.000} \\
  &\multicolumn{1}{@{~}c@{~}|@{~}}{BIM} & 0.998 & \textbf{0.999} & \textbf{0.978} & \textbf{1.000} & \textbf{0.995} & 0.937 & \textbf{1.000} \\
  &\multicolumn{1}{@{~}c@{~}|@{~}}{DF} & 0.993 & 0.998 & 0.944 & \textbf{1.000} & 0.987 & \textbf{0.939} & \textbf{1.000} \\
  \hline
  \hline
  \multirow{3}{*}{CIFAR-10} 
  &\multicolumn{1}{@{~}c@{~}|@{~}}{FGSM} & \textbf{0.992} & \textbf{1.000} & \textbf{1.000} & \textbf{1.000} & \textbf{0.994} & 0.911 & \textbf{1.000} \\
  &\multicolumn{1}{@{~}c@{~}|@{~}}{BIM} & \textbf{0.992} & \textbf{1.000} & \textbf{1.000} & \textbf{1.000} & 0.993 & 0.911 & \textbf{1.000} \\
  &\multicolumn{1}{@{~}c@{~}|@{~}}{DF} & 0.991 & 0.997 & 0.999 & 0.998 & \textbf{0.994} & \textbf{0.913} & 0.991 \\
  \hline\hline
  \multirow{3}{*}{ImageNet}
  &\multicolumn{1}{@{~}c@{~}|@{~}}{FGSM} & \textbf{0.926} & 0.991 & 0.999 & \textbf{0.994} & \textbf{0.999} & 0.981 & 0.997 \\
  &\multicolumn{1}{@{~}c@{~}|@{~}}{BIM} & 0.919 & \textbf{0.999} & \textbf{1.000} & 0.992 & 0.998 & \textbf{0.989} & \textbf{1.000} \\
  &\multicolumn{1}{@{~}c@{~}|@{~}}{DF} & 0.923 & 0.997 & 0.997 & 0.993 & 0.998 & 0.987 & 0.997 \\
  \hline
  \end{tabular}
  }
 \end{table}

\begin{figure}[t]
  \centering
    \hspace*{-2mm}\includegraphics[width=9.0cm]{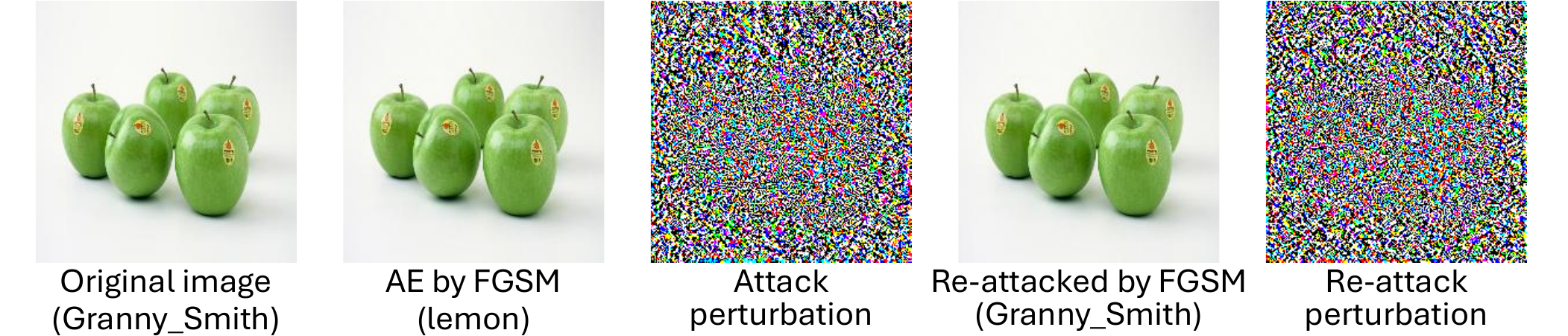}\\    
    \hspace*{-2mm}\includegraphics[width=9.0cm]{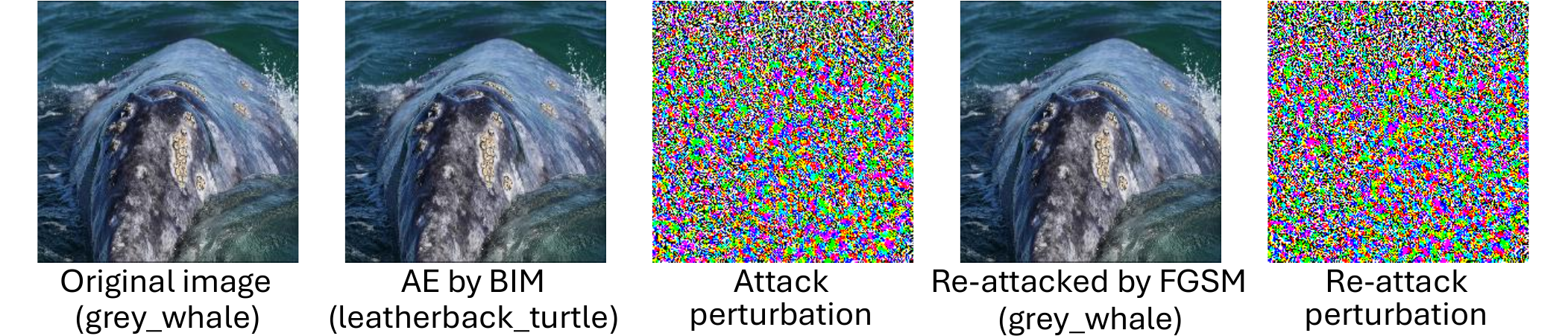}\\  
    \hspace*{-2mm}\includegraphics[width=9.0cm]{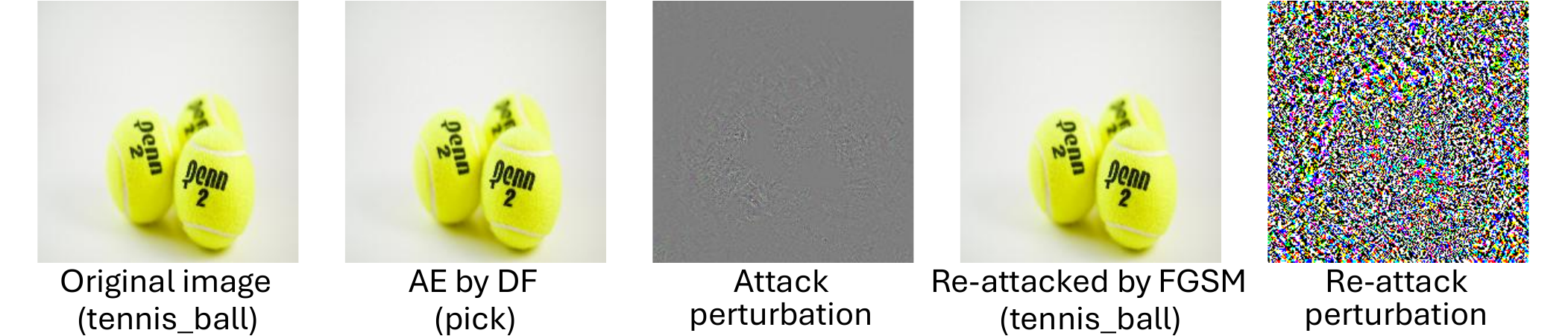}\\  
    \hspace*{-2mm}\includegraphics[width=9.0cm]{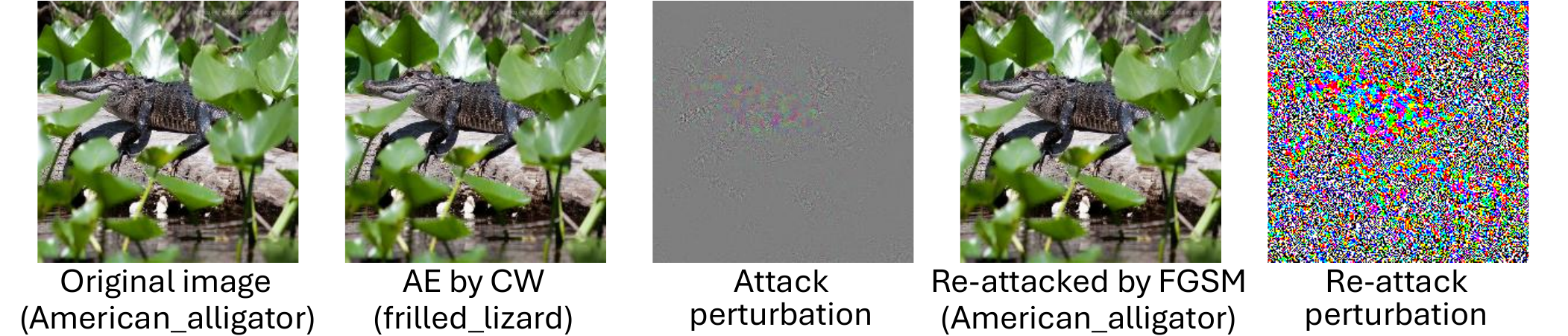}\\  
    \hspace*{-2mm}\includegraphics[width=9.0cm]{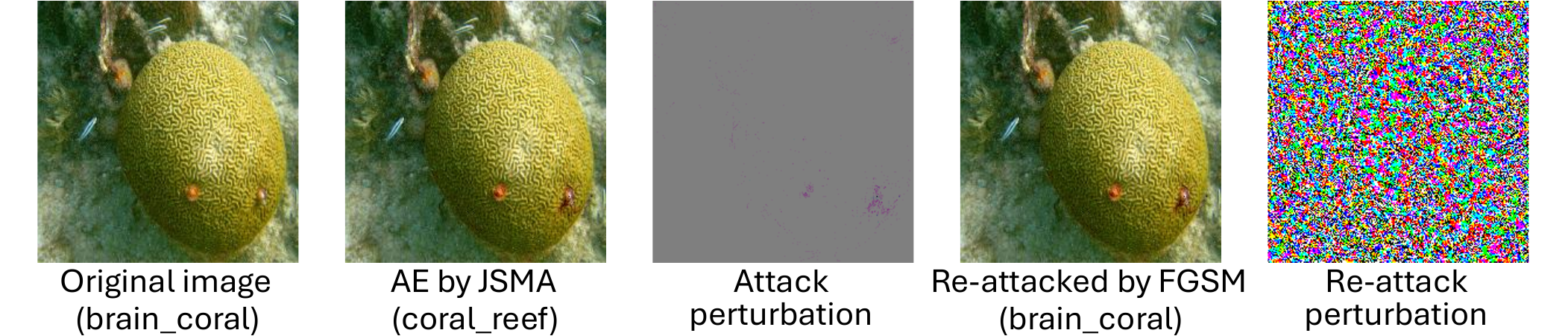}\\
    \hspace*{-2mm}\includegraphics[width=9.0cm]{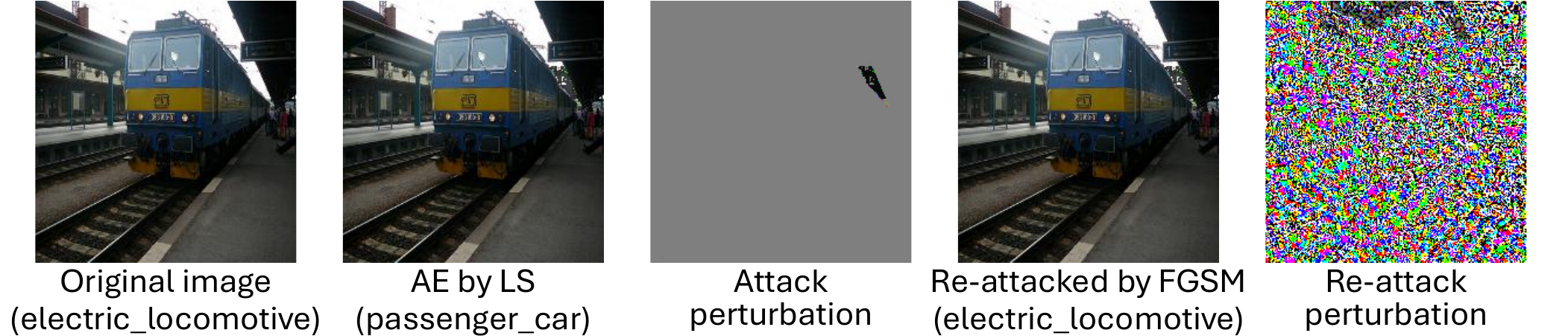}\\
    \hspace*{-2mm}\includegraphics[width=9.0cm]{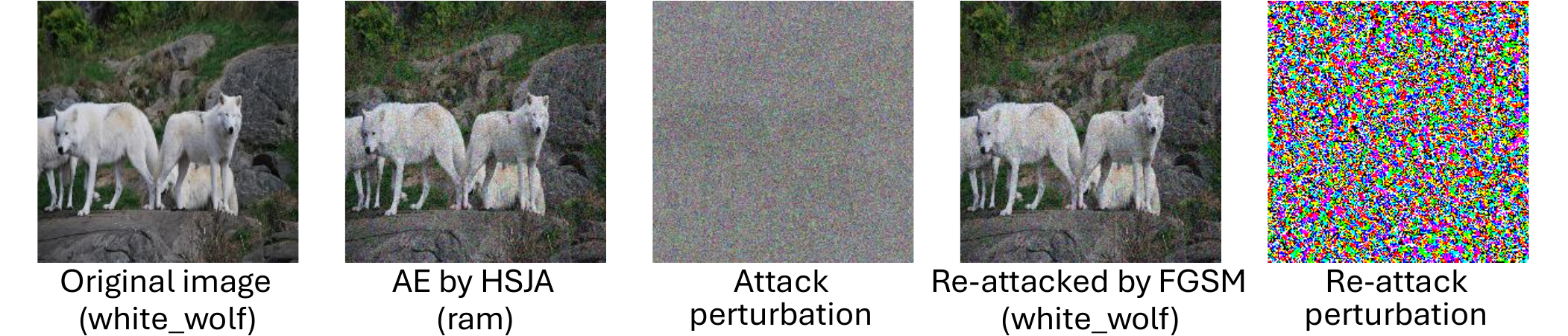}\\
    \caption{Example AEs rectified by our method re-attacking with FSGM in Experiment 1a.
        The labels in parentheses represent the recognition results by the classifier.
    }
    \label{fig:exp1_exam}
\end{figure}

\begin{table}[t]
  \centering
  \caption{Experiment 1a: Perturbation amount of AEs.}
  \label{tbl:attack_norm}
  {
  \begin{tabular}{@{~}c@{~}||@{~}c@{~}c@{~}c@{~}c@{~}c@{~}|@{~}c@{~}c@{~}}
  \hline
  \multicolumn{1}{@{~}c@{~}||@{~}}{\multirow{3}{*}{Dataset}} & 
  \multicolumn{7}{@{~}c@{~}}{\multirow{1}{*}{Attack method}} \\
  \cline{2-8}
  & 
  \multicolumn{5}{@{~}c@{~}|@{~}}{\multirow{1}{*}{White-box}} &
  \multicolumn{2}{@{~}c@{~}}{\multirow{1}{*}{Black-box}} \\
  \cline{2-8} 
  \multicolumn{1}{@{~}c@{~}||@{~}}{\multirow{2}{*}{}} & 
  \multicolumn{1}{@{~}c@{~}}{\multirow{1}{*}{FGSM}} & 
  \multicolumn{1}{@{~}c@{~}}{\multirow{1}{*}{BIM}} & 
  \multicolumn{1}{@{~}c@{~}}{\multirow{1}{*}{DF}} & 
  \multicolumn{1}{@{~}c@{~}}{\multirow{1}{*}{CW}} & 
  \multicolumn{1}{@{~}c@{~}|@{~}}{\multirow{1}{*}{JSMA}} & 
  \multicolumn{1}{@{~}c@{~}}{\multirow{1}{*}{LS}} & 
  \multicolumn{1}{@{~}c@{~}}{\multirow{1}{*}{HSJA}} \\ 
  \hline\hline
  \multicolumn{1}{@{~}c@{~}||@{~}}{\multirow{1}{*}{MNIST}} & 
  \multicolumn{1}{@{~}r@{~}}{\multirow{1}{*}{4.344}} & 
  \multicolumn{1}{@{~}r@{~}}{\multirow{1}{*}{2.487}} & 
  \multicolumn{1}{@{~}r@{~}}{\multirow{1}{*}{1.801}} & 
  \multicolumn{1}{@{~}r@{~}}{\multirow{1}{*}{1.398}} & 
  \multicolumn{1}{@{~}r@{~}|@{~}}{\multirow{1}{*}{2.964}} & 
  \multicolumn{1}{@{~}r@{~}}{\multirow{1}{*}{7.147}} & 
  \multicolumn{1}{@{~}r@{~}}{\multirow{1}{*}{1.591}} \\ 
  \hline
  \multicolumn{1}{@{~}c@{~}||@{~}}{\multirow{1}{*}{CIFAR-10}} & 
  \multicolumn{1}{@{~}r@{~}}{\multirow{1}{*}{1.139}} & 
  \multicolumn{1}{@{~}r@{~}}{\multirow{1}{*}{0.270}} & 
  \multicolumn{1}{@{~}r@{~}}{\multirow{1}{*}{0.196}} & 
  \multicolumn{1}{@{~}r@{~}}{\multirow{1}{*}{0.157}} & 
  \multicolumn{1}{@{~}r@{~}|@{~}}{\multirow{1}{*}{0.724}} & 
  \multicolumn{1}{@{~}r@{~}}{\multirow{1}{*}{5.756}} & 
  \multicolumn{1}{@{~}r@{~}}{\multirow{1}{*}{0.468}} \\ 
  \hline
  \multicolumn{1}{@{~}c@{~}||@{~}}{\multirow{1}{*}{ImageNet}} & 
  \multicolumn{1}{@{~}r@{~}}{\multirow{1}{*}{1.184}} & 
  \multicolumn{1}{@{~}r@{~}}{\multirow{1}{*}{0.235}} & 
  \multicolumn{1}{@{~}r@{~}}{\multirow{1}{*}{0.145}} & 
  \multicolumn{1}{@{~}r@{~}}{\multirow{1}{*}{0.154}} & 
  \multicolumn{1}{@{~}r@{~}|@{~}}{\multirow{1}{*}{1.243}} & 
  \multicolumn{1}{@{~}r@{~}}{\multirow{1}{*}{6.369}} & 
  \multicolumn{1}{@{~}r@{~}}{\multirow{1}{*}{23.074}} \\ 
  \hline
  \end{tabular}
  }
  \centering
  \caption{Experiment 1a: Perturbation amount of re-attack $(\times 10^{-3})$.}
  \label{tbl:re_attack_norm}
  {
  \begin{tabular}{@{~}c@{~}||@{~}c@{~}|@{~}c@{~}c@{~}c@{~}c@{~}c@{~}|@{~}c@{~}c@{~}}
  \hline
  \multirow{3}{*}{Dataset} &  
  \multirow{2}{*}{Re-attack} & 
  \multicolumn{7}{@{~}c@{~}}{\multirow{1}{*}{Attack method}} \\
  \cline{3-9}
  &
  \multirow{2}{*}{method}& 
  \multicolumn{5}{@{~}c@{~}|@{~}}{\multirow{1}{*}{White-box}} &
  \multicolumn{2}{@{~}c@{~}}{\multirow{1}{*}{Black-box}} \\
  \cline{3-9} 
  &
  \multicolumn{1}{@{~}c@{~}|@{~}}{} & 
  \multicolumn{1}{@{~}c@{~}}{\multirow{1}{*}{FGSM}} & 
  \multicolumn{1}{@{~}c@{~}}{\multirow{1}{*}{BIM}} & 
  \multicolumn{1}{@{~}c@{~}}{\multirow{1}{*}{DF}} & 
  \multicolumn{1}{@{~}c@{~}}{\multirow{1}{*}{CW}} & 
  \multirow{1}{*}{JSMA} & 
  \multicolumn{1}{@{~}c@{~}}{\multirow{1}{*}{LS}} & 
  \multicolumn{1}{@{~}c@{~}}{\multirow{1}{*}{HSJA}} \\ 
  \hline \hline
  \multirow{3}{*}{MNIST} &
  \multicolumn{1}{@{~}c@{~}|@{~}}{\multirow{1}{*}{FGSM}} & 
  \multicolumn{1}{@{~}r@{~}}{\multirow{1}{*}{5.8}} & 
  \multicolumn{1}{@{~}r@{~}}{\multirow{1}{*}{16.8}} & 
  \multicolumn{1}{@{~}r@{~}}{\multirow{1}{*}{210.7}} & 
  \multicolumn{1}{@{~}r@{~}}{\multirow{1}{*}{1.5}} & 
  \multicolumn{1}{@{~}r@{~}|@{~}}{\multirow{1}{*}{9.7}} & 
  \multicolumn{1}{@{~}r@{~}}{\multirow{1}{*}{153.7}} & 
  \multicolumn{1}{@{~}r@{~}}{\multirow{1}{*}{1.4}} \\ 
  &
  \multicolumn{1}{@{~}c@{~}|@{~}}{\multirow{1}{*}{BIM}} & 
  \multicolumn{1}{@{~}r@{~}}{\multirow{1}{*}{5.2}} & 
  \multicolumn{1}{@{~}r@{~}}{\multirow{1}{*}{15.6}} & 
  \multicolumn{1}{@{~}r@{~}}{\multirow{1}{*}{184.5}} & 
  \multicolumn{1}{@{~}r@{~}}{\multirow{1}{*}{0.6}} & 
  \multicolumn{1}{@{~}r@{~}|@{~}}{\multirow{1}{*}{8.9}} & 
  \multicolumn{1}{@{~}r@{~}}{\multirow{1}{*}{141.8}} & 
  \multicolumn{1}{@{~}r@{~}}{\multirow{1}{*}{0.1}} \\ 
  &
  \multicolumn{1}{@{~}c@{~}|@{~}}{\multirow{1}{*}{DF}} & 
  \multicolumn{1}{@{~}r@{~}}{\multirow{1}{*}{110.7}} & 
  \multicolumn{1}{@{~}r@{~}}{\multirow{1}{*}{13.7}} & 
  \multicolumn{1}{@{~}r@{~}}{\multirow{1}{*}{317.4}} & 
  \multicolumn{1}{@{~}r@{~}}{\multirow{1}{*}{1.0}} & 
  \multicolumn{1}{@{~}r@{~}|@{~}}{\multirow{1}{*}{113.8}} & 
  \multicolumn{1}{@{~}r@{~}}{\multirow{1}{*}{169.4}} & 
  \multicolumn{1}{@{~}r@{~}}{\multirow{1}{*}{7.6}} \\ 
  \hline \hline
  \multirow{3}{*}{CIFAR-10} &
  \multicolumn{1}{@{~}c@{~}|@{~}}{\multirow{1}{*}{FGSM}} & 
  \multicolumn{1}{@{~}r@{~}}{\multirow{1}{*}{6.1}} & 
  \multicolumn{1}{@{~}r@{~}}{\multirow{1}{*}{3.2}} & 
  \multicolumn{1}{@{~}r@{~}}{\multirow{1}{*}{6.9}} &  
  \multicolumn{1}{@{~}r@{~}}{\multirow{1}{*}{2.8}} & 
  \multicolumn{1}{@{~}r@{~}|@{~}}{\multirow{1}{*}{8.0}} & 
  \multicolumn{1}{@{~}r@{~}}{\multirow{1}{*}{35.2}} &
  \multicolumn{1}{@{~}r@{~}}{\multirow{1}{*}{2.8}} \\ 
  &
  \multicolumn{1}{@{~}c@{~}|@{~}}{\multirow{1}{*}{BIM}} & 
  \multicolumn{1}{@{~}r@{~}}{\multirow{1}{*}{4.6}} & 
  \multicolumn{1}{@{~}r@{~}}{\multirow{1}{*}{0.8}} & 
  \multicolumn{1}{@{~}r@{~}}{\multirow{1}{*}{4.7}} & 
  \multicolumn{1}{@{~}r@{~}}{\multirow{1}{*}{0.5}} & 
  \multicolumn{1}{@{~}r@{~}|@{~}}{\multirow{1}{*}{6.4}} & 
  \multicolumn{1}{@{~}r@{~}}{\multirow{1}{*}{30.5}} &  
  \multicolumn{1}{@{~}r@{~}}{\multirow{1}{*}{0.1}} \\ 
  &
  \multicolumn{1}{@{~}c@{~}|@{~}}{\multirow{1}{*}{DF}} & 
  \multicolumn{1}{@{~}r@{~}}{\multirow{1}{*}{3.2}} & 
  \multicolumn{1}{@{~}r@{~}}{\multirow{1}{*}{0.5}} & 
  \multicolumn{1}{@{~}r@{~}}{\multirow{1}{*}{3.0}} & 
  \multicolumn{1}{@{~}r@{~}}{\multirow{1}{*}{0.3}} & 
  \multicolumn{1}{@{~}r@{~}|@{~}}{\multirow{1}{*}{4.1}} & 
  \multicolumn{1}{@{~}r@{~}}{\multirow{1}{*}{19.4}} & 
  \multicolumn{1}{@{~}r@{~}}{\multirow{1}{*}{0.6}} \\ 
  \hline \hline
  \multirow{3}{*}{ImageNet} &
  \multicolumn{1}{@{~}c@{~}|@{~}}{\multirow{1}{*}{FGSM}} & 
  \multicolumn{1}{@{~}r@{~}}{\multirow{1}{*}{20.5}} & 
  \multicolumn{1}{@{~}r@{~}}{\multirow{1}{*}{20.6}} & 
  \multicolumn{1}{@{~}r@{~}}{\multirow{1}{*}{19.4}} & 
  \multicolumn{1}{@{~}r@{~}}{\multirow{1}{*}{23.7}} & 
  \multicolumn{1}{@{~}r@{~}|@{~}}{\multirow{1}{*}{19.3}} & 
  \multicolumn{1}{@{~}r@{~}}{\multirow{1}{*}{19.4}} & 
  \multicolumn{1}{@{~}r@{~}}{\multirow{1}{*}{19.4}} \\ 
  &
  \multicolumn{1}{@{~}c@{~}|@{~}}{\multirow{1}{*}{BIM}} & 
  \multicolumn{1}{@{~}r@{~}}{\multirow{1}{*}{8.4}} & 
  \multicolumn{1}{@{~}r@{~}}{\multirow{1}{*}{1.0}} & 
  \multicolumn{1}{@{~}r@{~}}{\multirow{1}{*}{0.5}} & 
  \multicolumn{1}{@{~}r@{~}}{\multirow{1}{*}{8.0}} & 
  \multicolumn{1}{@{~}r@{~}|@{~}}{\multirow{1}{*}{1.4}} & 
  \multicolumn{1}{@{~}r@{~}}{\multirow{1}{*}{2.6}} & 
  \multicolumn{1}{@{~}r@{~}}{\multirow{1}{*}{0.1}} \\ 
  &
  \multicolumn{1}{@{~}c@{~}|@{~}}{\multirow{1}{*}{DF}} & 
  \multicolumn{1}{@{~}r@{~}}{\multirow{1}{*}{4.4}} & 
  \multicolumn{1}{@{~}r@{~}}{\multirow{1}{*}{0.8}} & 
  \multicolumn{1}{@{~}r@{~}}{\multirow{1}{*}{0.4}} & 
  \multicolumn{1}{@{~}r@{~}}{\multirow{1}{*}{4.6}} & 
  \multicolumn{1}{@{~}r@{~}|@{~}}{\multirow{1}{*}{0.9}} &  
  \multicolumn{1}{@{~}r@{~}}{\multirow{1}{*}{1.4}} & 
  \multicolumn{1}{@{~}r@{~}}{\multirow{1}{*}{0.1}} \\ 
  \hline
  \end{tabular}
  }
  \centering
  \caption{Experiment 1a: Cosine similarities between attack perturbations and inverses of re-attack ones,
      which indicate the appropriateness of re-attack direction.
}
  \label{tbl:cos_sim}
  {
  \begin{tabular}{@{~}c@{~}||@{~}c@{~}|@{~}c@{~}c@{~}c@{~}c@{~}c@{~}|@{~}c@{~}c@{~}}
  \hline
  \multirow{3}{*}{Dataset} &  
  \multirow{2}{*}{Re-attack} & 
  \multicolumn{7}{@{~}c@{~}}{\multirow{1}{*}{Attack method}} \\
  \cline{3-9}
  &
  \multirow{2}{*}{method}& 
  \multicolumn{5}{@{~}c@{~}|@{~}}{\multirow{1}{*}{White-box}} &
  \multicolumn{2}{@{~}c@{~}}{\multirow{1}{*}{Black-box}} \\
  \cline{3-9} 
  &
  \multicolumn{1}{@{~}c@{~}|@{~}}{} & 
  \multicolumn{1}{@{~}c@{~}}{\multirow{1}{*}{FGSM}} & 
  \multicolumn{1}{@{~}c@{~}}{\multirow{1}{*}{BIM}} & 
  \multicolumn{1}{@{~}c@{~}}{\multirow{1}{*}{DF}} & 
  \multicolumn{1}{@{~}c@{~}}{\multirow{1}{*}{CW}} & 
  \multirow{1}{*}{JSMA} & 
  \multicolumn{1}{@{~}c@{~}}{\multirow{1}{*}{LS}} & 
  \multicolumn{1}{@{~}c@{~}}{\multirow{1}{*}{HSJA}} \\ 
  \hline \hline
  \multirow{3}{*}{MNIST} &
  \multicolumn{1}{@{~}c@{~}|@{~}}{\multirow{1}{*}{FGSM}} & 
  \multicolumn{1}{@{~}c@{~}}{\multirow{1}{*}{0.232}} &  
  \multicolumn{1}{@{~}c@{~}}{\multirow{1}{*}{0.580}} & 
  \multicolumn{1}{@{~}c@{~}}{\multirow{1}{*}{0.336}} & 
  \multicolumn{1}{@{~}c@{~}}{\multirow{1}{*}{0.412}} & 
  \multicolumn{1}{@{~}c@{~}|@{~}}{\multirow{1}{*}{0.202}} & 
  \multicolumn{1}{@{~}c@{~}}{\multirow{1}{*}{0.058}} &
  \multicolumn{1}{@{~}c@{~}}{\multirow{1}{*}{0.384}} \\ 
  &
  \multicolumn{1}{@{~}c@{~}|@{~}}{\multirow{1}{*}{BIM}} & 
  \multicolumn{1}{@{~}c@{~}}{\multirow{1}{*}{0.232}} & 
  \multicolumn{1}{@{~}c@{~}}{\multirow{1}{*}{0.583}} & 
  \multicolumn{1}{@{~}c@{~}}{\multirow{1}{*}{0.359}} & 
  \multicolumn{1}{@{~}c@{~}}{\multirow{1}{*}{0.408}} & 
  \multicolumn{1}{@{~}c@{~}|@{~}}{\multirow{1}{*}{0.204}} & 
  \multicolumn{1}{@{~}c@{~}}{\multirow{1}{*}{0.061}} & 
  \multicolumn{1}{@{~}c@{~}}{\multirow{1}{*}{0.384}} \\ 
  &
  \multicolumn{1}{@{~}c@{~}|@{~}}{\multirow{1}{*}{DF}} & 
  \multicolumn{1}{@{~}c@{~}}{\multirow{1}{*}{0.266}} & 
  \multicolumn{1}{@{~}c@{~}}{\multirow{1}{*}{0.503}} & 
  \multicolumn{1}{@{~}c@{~}}{\multirow{1}{*}{0.438}} & 
  \multicolumn{1}{@{~}c@{~}}{\multirow{1}{*}{0.887}} & 
  \multicolumn{1}{@{~}c@{~}|@{~}}{\multirow{1}{*}{0.490}} & 
  \multicolumn{1}{@{~}c@{~}}{\multirow{1}{*}{0.117}} &
  \multicolumn{1}{@{~}c@{~}}{\multirow{1}{*}{0.827}} \\ 
  \hline \hline
  \multirow{3}{*}{CIFAR-10} &
  \multicolumn{1}{@{~}c@{~}|@{~}}{\multirow{1}{*}{FGSM}} & 
  \multicolumn{1}{@{~}c@{~}}{\multirow{1}{*}{0.448}} & 
  \multicolumn{1}{@{~}c@{~}}{\multirow{1}{*}{0.834}} & 
  \multicolumn{1}{@{~}c@{~}}{\multirow{1}{*}{0.478}} & 
  \multicolumn{1}{@{~}c@{~}}{\multirow{1}{*}{0.568}} & 
  \multicolumn{1}{@{~}c@{~}|@{~}}{\multirow{1}{*}{0.044}} & 
  \multicolumn{1}{@{~}c@{~}}{\multirow{1}{*}{0.001}} & 
  \multicolumn{1}{@{~}c@{~}}{\multirow{1}{*}{0.210}} \\ 
  &
  \multicolumn{1}{@{~}c@{~}|@{~}}{\multirow{1}{*}{BIM}} & 
  \multicolumn{1}{@{~}c@{~}}{\multirow{1}{*}{0.451}} & 
  \multicolumn{1}{@{~}c@{~}}{\multirow{1}{*}{0.833}} & 
  \multicolumn{1}{@{~}c@{~}}{\multirow{1}{*}{0.479}} & 
  \multicolumn{1}{@{~}c@{~}}{\multirow{1}{*}{0.568}} & 
  \multicolumn{1}{@{~}c@{~}|@{~}}{\multirow{1}{*}{0.045}} & 
  \multicolumn{1}{@{~}c@{~}}{\multirow{1}{*}{0.001}} & 
  \multicolumn{1}{@{~}c@{~}}{\multirow{1}{*}{0.208}} \\ 
  &
  \multicolumn{1}{@{~}c@{~}|@{~}}{\multirow{1}{*}{DF}} & 
  \multicolumn{1}{@{~}c@{~}}{\multirow{1}{*}{0.367}} & 
  \multicolumn{1}{@{~}c@{~}}{\multirow{1}{*}{0.585}} & 
  \multicolumn{1}{@{~}c@{~}}{\multirow{1}{*}{0.778}} & 
  \multicolumn{1}{@{~}c@{~}}{\multirow{1}{*}{0.978}} & 
  \multicolumn{1}{@{~}c@{~}|@{~}}{\multirow{1}{*}{0.149}} & 
  \multicolumn{1}{@{~}c@{~}}{\multirow{1}{*}{0.008}} & 
  \multicolumn{1}{@{~}c@{~}}{\multirow{1}{*}{0.342}} \\ 
  \hline \hline
  \multirow{3}{*}{ImageNet} &
  \multicolumn{1}{@{~}c@{~}|@{~}}{\multirow{1}{*}{FGSM}} & 
  \multicolumn{1}{@{~}c@{~}}{\multirow{1}{*}{0.316}} &  
  \multicolumn{1}{@{~}c@{~}}{\multirow{1}{*}{0.618}} & 
  \multicolumn{1}{@{~}c@{~}}{\multirow{1}{*}{0.446}} & 
  \multicolumn{1}{@{~}c@{~}}{\multirow{1}{*}{0.469}} & 
  \multicolumn{1}{@{~}c@{~}|@{~}}{\multirow{1}{*}{0.006}} & 
  \multicolumn{1}{@{~}c@{~}}{\multirow{1}{*}{0.001}} & 
  \multicolumn{1}{@{~}c@{~}}{\multirow{1}{*}{0.005}} \\ 
  &
  \multicolumn{1}{@{~}c@{~}|@{~}}{\multirow{1}{*}{BIM}} & 
  \multicolumn{1}{@{~}c@{~}}{\multirow{1}{*}{0.324}} & 
  \multicolumn{1}{@{~}c@{~}}{\multirow{1}{*}{0.619}} & 
  \multicolumn{1}{@{~}c@{~}}{\multirow{1}{*}{0.446}} & 
  \multicolumn{1}{@{~}c@{~}}{\multirow{1}{*}{0.476}} & 
  \multicolumn{1}{@{~}c@{~}|@{~}}{\multirow{1}{*}{0.006}} & 
  \multicolumn{1}{@{~}c@{~}}{\multirow{1}{*}{0.001}} & 
  \multicolumn{1}{@{~}c@{~}}{\multirow{1}{*}{0.005}} \\ 
  &
  \multicolumn{1}{@{~}c@{~}|@{~}}{\multirow{1}{*}{DF}} & 
  \multicolumn{1}{@{~}c@{~}}{\multirow{1}{*}{0.319}} & 
  \multicolumn{1}{@{~}c@{~}}{\multirow{1}{*}{0.500}} & 
  \multicolumn{1}{@{~}c@{~}}{\multirow{1}{*}{0.780}} & 
  \multicolumn{1}{@{~}c@{~}}{\multirow{1}{*}{0.842}} & 
  \multicolumn{1}{@{~}c@{~}|@{~}}{\multirow{1}{*}{0.029}} & 
  \multicolumn{1}{@{~}c@{~}}{\multirow{1}{*}{0.001}} & 
  \multicolumn{1}{@{~}c@{~}}{\multirow{1}{*}{0.011}} \\ 
  \hline
  \end{tabular}
  }
\end{table}

\begin{table*}[t]
  \centering
  \caption{Experiment 1b: Success rates of rectification (targeted attack).}
  \label{tbl:target_attack}
  \begin{minipage}[t]{0.315\textwidth}
    \centering
  {(a) Success rates in MNIST}
  {
  \begin{tabular}{@{~}c@{~}|@{~}c@{~}|@{~}c@{~~}c@{~~}c@{~~}c@{~}}
  \hline
  \multicolumn{1}{@{~}c@{~}|@{~}}{Attack} & \multicolumn{1}{@{~}c@{~}|@{~}}{Re-attack}  & \multicolumn{4}{@{~}c@{~}}{Target label} \\ \cline{3-6} 
  \multicolumn{1}{@{~}c@{~}|@{~}}{method} & \multicolumn{1}{@{~}c@{~}|@{~}}{method} & \multicolumn{1}{@{~}c@{~}}{Top-2} & \multicolumn{1}{@{~}c@{~}}{Top-3} & \multicolumn{1}{@{~}c@{~}}{Top-4} & \multicolumn{1}{@{~}c@{~}}{Top-5} \\
  \hline\hline
  \multicolumn{1}{@{~}c@{~}|@{~}} {\multirow{3}{*}{FGSM}}
  & \multicolumn{1}{@{~}c@{~}|@{~}}{FGSM} & \textbf{0.990} & \textbf{0.299} & \textbf{0.158} & \textbf{0.172} \\
  & \multicolumn{1}{@{~}c@{~}|@{~}}{BIM} & \textbf{0.990} & 0.298 & \textbf{0.158} & \textbf{0.172} \\
  & \multicolumn{1}{@{~}c@{~}|@{~}}{DF} & 0.979 & 0.297 & 0.157 & \textbf{0.172} \\
  \hline 
  \multicolumn{1}{@{~}c@{~}|@{~}} {\multirow{3}{*}{BIM}}
  & \multicolumn{1}{@{~}c@{~}|@{~}}{FGSM} & 0.992 & 0.966 & 0.934 & 0.871 \\
  & \multicolumn{1}{@{~}c@{~}|@{~}}{BIM} & 0.992 & 0.966 & 0.932 & 0.871 \\
  & \multicolumn{1}{@{~}c@{~}|@{~}}{DF} & \textbf{0.994} & \textbf{0.970} & \textbf{0.940} & \textbf{0.878} \\
  \hline
  \multicolumn{1}{@{~}c@{~}|@{~}} {\multirow{3}{*}{CW}}
  & \multicolumn{1}{@{~}c@{~}|@{~}}{FGSM} & 0.993 & \textbf{0.982} & 0.961 & \textbf{0.948} \\
  & \multicolumn{1}{@{~}c@{~}|@{~}}{BIM} & 0.993 & 0.979 & 0.957 & 0.941 \\
  & \multicolumn{1}{@{~}c@{~}|@{~}}{DF} & \textbf{0.995} & 0.980 & \textbf{0.962} & 0.937 \\
  \hline
  \multicolumn{1}{@{~}c@{~}|@{~}} {\multirow{3}{*}{JSMA}}
  & \multicolumn{1}{@{~}c@{~}|@{~}}{FGSM} & \textbf{0.988} & \textbf{0.955} & 0.936 & \textbf{0.870} \\
  & \multicolumn{1}{@{~}c@{~}|@{~}}{BIM} & \textbf{0.988} & \textbf{0.955} & \textbf{0.937} & \textbf{0.870} \\
  & \multicolumn{1}{@{~}c@{~}|@{~}}{DF} & 0.976 & 0.937 & 0.916 & 0.859 \\
  \hline
  \multicolumn{1}{@{~}c@{~}|@{~}} {\multirow{3}{*}{HSJA}}
  & \multicolumn{1}{@{~}c@{~}|@{~}}{FGSM} & 0.994 & \textbf{0.971} & 0.952 & \textbf{0.928} \\
  & \multicolumn{1}{@{~}c@{~}|@{~}}{BIM} & 0.998 & 0.968 & 0.956 & 0.926 \\
  & \multicolumn{1}{@{~}c@{~}|@{~}}{DF} & \textbf{0.999} & 0.969 & \textbf{0.958} & 0.926 \\
  \hline
  \end{tabular}
  }
  \end{minipage}~~~
  \begin{minipage}[t]{0.315\textwidth}
    \centering
  {(b) Success rates in CIFAR-10} 
  {
  \begin{tabular}{@{~}c@{~}|@{~}c@{~}|@{~}c@{~~}c@{~~}c@{~~}c@{~}}
  \hline
  \multicolumn{1}{@{~}c@{~}|@{~}}{Attack} & \multicolumn{1}{@{~}c@{~}|@{~}}{Re-attack}  & \multicolumn{4}{@{~}c@{~}}{Target label}                                                                                                                                                      \\ \cline{3-6} 
  \multicolumn{1}{@{~}c@{~}|@{~}}{method} & \multicolumn{1}{@{~}c@{~}|@{~}}{method} & \multicolumn{1}{@{~}c@{~}}{Top-2} & \multicolumn{1}{@{~}c@{~}}{Top-3} & \multicolumn{1}{@{~}c@{~}}{Top-4} & \multicolumn{1}{@{~}c@{~}}{Top-5} \\
  \hline\hline
  \multicolumn{1}{@{~}c@{~}|@{~}} {\multirow{3}{*}{FGSM}} & \multicolumn{1}{@{~}c@{~}|@{~}}{FGSM} & 0.957 & 0.270 & 0.152 & \textbf{0.122} \\
  & \multicolumn{1}{@{~}c@{~}|@{~}}{BIM} & 0.957 & \textbf{0.271} & 0.152 & \textbf{0.122} \\
  & \multicolumn{1}{@{~}c@{~}|@{~}}{DF} & \textbf{0.966} & \textbf{0.271} & \textbf{0.153} & \textbf{0.122} \\
  \hline 
  \multicolumn{1}{@{~}c@{~}|@{~}} {\multirow{3}{*}{BIM}} & \multicolumn{1}{@{~}c@{~}|@{~}}{FGSM} & 0.997 & \textbf{1.000} & \textbf{0.908} & \textbf{0.856} \\
  & \multicolumn{1}{@{~}c@{~}|@{~}}{BIM} & \textbf{0.998} & 0.937 & 0.898 & 0.848 \\
  & \multicolumn{1}{@{~}c@{~}|@{~}}{DF} & 0.994 & 0.936 & 0.901 & 0.847 \\
  \hline
  \multicolumn{1}{@{~}c@{~}|@{~}} {\multirow{3}{*}{CW}} & \multicolumn{1}{@{~}c@{~}|@{~}}{FGSM} & 0.997 & \textbf{0.945} & \textbf{0.915} & \textbf{0.877} \\
  & \multicolumn{1}{@{~}c@{~}|@{~}}{BIM} & \textbf{0.998} & 0.937 & 0.898 & 0.848 \\
  & \multicolumn{1}{@{~}c@{~}|@{~}}{DF} & 0.997 & 0.934 & 0.898 & 0.845 \\
  \hline
  \multicolumn{1}{@{~}c@{~}|@{~}} {\multirow{3}{*}{JSMA}} & \multicolumn{1}{@{~}c@{~}|@{~}}{FGSM} & 0.993 & \textbf{0.896} & 0.826 & \textbf{0.767} \\
  & \multicolumn{1}{@{~}c@{~}|@{~}}{BIM} & 0.993 & 0.895 & \textbf{0.827} & 0.766 \\
  & \multicolumn{1}{@{~}c@{~}|@{~}}{DF} & \textbf{0.995} & 0.891 & 0.823 & 0.758 \\
  \hline
  \multicolumn{1}{@{~}c@{~}|@{~}} {\multirow{3}{*}{HSJA}} & \multicolumn{1}{@{~}c@{~}|@{~}}{FGSM} & 0.997 & \textbf{0.939} & \textbf{0.906} & \textbf{0.863} \\
  & \multicolumn{1}{@{~}c@{~}|@{~}}{BIM} & \textbf{0.999} & 0.927 & 0.889 & 0.838 \\
  & \multicolumn{1}{@{~}c@{~}|@{~}}{DF} & 0.992 & 0.920 & 0.882 & 0.828 \\
  \hline
  \end{tabular}
  }
  \end{minipage}~~~
  \begin{minipage}[t]{0.315\textwidth}
  \centering
  {(c) Success rates in ImageNet}
  {
  \begin{tabular}{@{~}c@{~}|@{~}c@{~}|@{~}c@{~~}c@{~~}c@{~~}c@{~}}
  \hline
  \multicolumn{1}{@{~}c@{~}|@{~}}{Attack} & \multicolumn{1}{@{~}c@{~}|@{~}}{Re-attack}  & \multicolumn{4}{@{~}c@{~}}{Target label}                                                                                                                                                      \\ \cline{3-6} 
  \multicolumn{1}{@{~}c@{~}|@{~}}{Method} & \multicolumn{1}{@{~}c@{~}|@{~}}{Method} & \multicolumn{1}{@{~}c@{~}}{Top-2} & \multicolumn{1}{@{~}c@{~}}{Top-3} & \multicolumn{1}{@{~}c@{~}}{Top-4} & \multicolumn{1}{@{~}c@{~}}{Top-5} \\
  \hline\hline
  \multicolumn{1}{@{~}c@{~}|@{~}} {\multirow{3}{*}{FGSM}} & \multicolumn{1}{@{~}c@{~}|@{~}}{FGSM} & \textbf{0.917} & \textbf{0.390} & \textbf{0.251} & \textbf{0.156} \\
  & \multicolumn{1}{@{~}c@{~}|@{~}}{BIM} & 0.915 & 0.388 & 0.242 & 0.142 \\
  & \multicolumn{1}{@{~}c@{~}|@{~}}{DF} & 0.915 & 0.385 & 0.241 & 0.144 \\
  \hline 
  \multicolumn{1}{@{~}c@{~}|@{~}} {\multirow{3}{*}{BIM}} & \multicolumn{1}{@{~}c@{~}|@{~}}{FGSM} & 0.997 & \textbf{0.952} & \textbf{0.915} & \textbf{0.891} \\
  & \multicolumn{1}{@{~}c@{~}|@{~}}{BIM} & 0.997 & 0.924 & 0.895 & 0.857 \\
  & \multicolumn{1}{@{~}c@{~}|@{~}}{DF} & \textbf{0.998} & 0.923 & 0.892 & 0.853 \\
  \hline
  \multicolumn{1}{@{~}c@{~}|@{~}} {\multirow{3}{*}{CW}} & \multicolumn{1}{@{~}c@{~}|@{~}}{FGSM} & 0.972 & 0.891 & \textbf{0.849} & \textbf{0.833} \\
  & \multicolumn{1}{@{~}c@{~}|@{~}}{BIM} & 0.977 & 0.878 & 0.830 & 0.806 \\
  & \multicolumn{1}{@{~}c@{~}|@{~}}{DF} & \textbf{0.987} & \textbf{0.892} & 0.831 & 0.813 \\
  \hline
  \multicolumn{1}{@{~}c@{~}|@{~}} {\multirow{3}{*}{JSMA}} & \multicolumn{1}{@{~}c@{~}|@{~}}{FGSM} & \textbf{0.997} & \textbf{0.914} & \textbf{0.877} & \textbf{0.852} \\
  & \multicolumn{1}{@{~}c@{~}|@{~}}{BIM} & 0.993 & 0.898 & 0.853 & 0.832 \\
  & \multicolumn{1}{@{~}c@{~}|@{~}}{DF} & 0.993 & 0.898 & 0.851 & 0.829 \\
  \hline
  \multicolumn{1}{@{~}c@{~}|@{~}} {\multirow{3}{*}{HSJA}} & \multicolumn{1}{@{~}c@{~}|@{~}}{FGSM} & 0.728 & 0.602 & 0.565 & 0.517 \\
  & \multicolumn{1}{@{~}c@{~}|@{~}}{BIM} & \textbf{0.820} & \textbf{0.692} & \textbf{0.637} & \textbf{0.568} \\
  & \multicolumn{1}{@{~}c@{~}|@{~}}{DF} & 0.818 & 0.687 & 0.635 & 0.567 \\
  \hline
  \end{tabular}
  }
  \end{minipage}
\end{table*}

\begin{table}[t]
  \centering
  \caption{Experiment 1b: Perturbation amount of AEs generated by targeted attack methods.}
  \label{tbl:target_norm}
  {
    
  \begin{tabular}{@{~}c@{~}||@{~}c@{~}|@{~}r@{~~}r@{~~}r@{~~}r@{~}}
  \hline
  \multicolumn{1}{@{~}c@{~}||@{~}}{\multirow{2}{*}{Dataset}} & \multicolumn{1}{@{~}c@{~}|@{~}}{Attack}  & \multicolumn{4}{@{~}c@{~}}{Target label}                                                                                                                                                      \\ \cline{3-6} 
  \multicolumn{1}{@{~}c@{~}||@{~}}{} & \multicolumn{1}{@{~}c@{~}|@{~}}{method} & \multicolumn{1}{@{~}c@{~}}{Top-2} & \multicolumn{1}{@{~}c@{~}}{Top-3} & \multicolumn{1}{@{~}c@{~}}{Top-4} & \multicolumn{1}{@{~}c@{~}}{Top-5} \\
  \hline\hline
  \multicolumn{1}{@{~}c@{~}||@{~}} {\multirow{5}{*}{MNIST}} & \multicolumn{1}{@{~}c@{~}|@{~}}{FGSM} & 
  4.133 & 10.133 & 12.125 & 12.626 \\
  \multicolumn{1}{@{~}l@{~}||@{~}} {} & \multicolumn{1}{@{~}c@{~}|@{~}}{BIM} & 
  2.499 & 2.882 & 3.147 & 3.394 \\
  \multicolumn{1}{@{~}l@{~}||@{~}} {} & \multicolumn{1}{@{~}c@{~}|@{~}}{CW} & 
  1.417 & 1.649 & 1.813 & 1.990 \\
  \multicolumn{1}{@{~}l@{~}||@{~}} {} & \multicolumn{1}{@{~}c@{~}|@{~}}{JSMA} & 
  2.707& 3.007& 3.324 & 3.497 \\
  \multicolumn{1}{@{~}l@{~}||@{~}} {} & \multicolumn{1}{@{~}c@{~}|@{~}}{HSJA} & 
  1.560 & 1.813 & 1.992 & 2.148 \\
  \hline\hline
  \multicolumn{1}{@{~}c@{~}||@{~}} {\multirow{5}{*}{CIFAR-10}} & \multicolumn{1}{@{~}c@{~}|@{~}}{FGSM} & 
  1.064 & 6.075 & 8.547 & 9.179 \\
  \multicolumn{1}{@{~}l@{~}||@{~}} {} & \multicolumn{1}{@{~}c@{~}|@{~}}{BIM} & 
  0.266 & 0.348 & 0.399 & 0.444 \\
  \multicolumn{1}{@{~}l@{~}||@{~}} {} & \multicolumn{1}{@{~}c@{~}|@{~}}{CW} & 
  0.161 & 0.211& 0.242 & 0.268 \\
  \multicolumn{1}{@{~}l@{~}||@{~}} {} & \multicolumn{1}{@{~}c@{~}|@{~}}{JSMA} & 
  0.699 & 0.871 & 0.974 & 1.061 \\
  \multicolumn{1}{@{~}l@{~}||@{~}} {} & \multicolumn{1}{@{~}c@{~}|@{~}}{HSJA} & 
  0.432 & 0.546 & 0.607 & 0.655 \\
  \hline\hline	
  \multicolumn{1}{@{~}c@{~}||@{~}} {\multirow{5}{*}{ImageNet}} & \multicolumn{1}{@{~}c@{~}|@{~}}{FGSM} & 
  2.354 & 12.462 & 19.405 & 23.596 \\
  \multicolumn{1}{@{~}l@{~}||@{~}} {} & \multicolumn{1}{@{~}c@{~}|@{~}}{BIM} & 
  0.226 & 0.286 & 0.309 & 0.334\\
  \multicolumn{1}{@{~}l@{~}||@{~}} {} & \multicolumn{1}{@{~}c@{~}|@{~}}{CW} & 
  0.160 & 0.186 & 0.190 & 0.201 \\
  \multicolumn{1}{@{~}l@{~}||@{~}} {} & \multicolumn{1}{@{~}c@{~}|@{~}}{JSMA} & 
  1.344 & 1.620 & 1.737& 1.810 \\
  \multicolumn{1}{@{~}l@{~}||@{~}} {} & \multicolumn{1}{@{~}c@{~}|@{~}}{HSJA} & 
  39.999 & 47.810 & 50.421 & 53.445 \\
  \hline
  \end{tabular}

  }
\end{table}

\subsubsection{Results on rectification performance}

Table~\ref{tbl:attack_comparison} presents the rectification success
rates of the proposed method with three re-attack methods: FGSM, BIM, and
DF.
It
demonstrates that the proposed method
can be effectively applied across a wide range of datasets and attack
methods, successfully rectifying more than 90\% AEs created using all seven attack
methods on all three datasets.
Notably, the proposed method accurately estimated the correct labels
even when a re-attack method different from that used for
the initial attack was employed.
This validates the core concept of the proposed method, demonstrating that
rectification is achievable by leveraging the close distance between AEs and
decision boundary.

Furthermore, our method demonstrated the capability to effectively
rectify AEs generated by black-box attacks such as LS and HSJA,
which do not rely on the loss function gradients in DNNs.
The success rates for LS were lower than those for white-box attack
methods, attributed to its lack of a mechanism for discovering AEs on
the discrimination boundary edge.
However, our method maintains around 91\% success rates in the worst cases,
which constitute sufficiently high success rates as will be evident from the comparison 
with a previous method detailed in Section~\ref{ssec:vs_rsv}.

Figure~\ref{fig:exp1_exam} depicts the application results of our
method through  examples of attempted rectifications via re-attack
on ImageNet.
Each row contains five images, from left to right: an input image, an AE, AE's
perturbation, a rectified AE, and a re-attack perturbation.
For example, in the top row, the first example showcases a re-attack
with FGSM in response to an initial FGSM attack, demonstrating
successful label correction.
The sixth example
depicts the result of a re-attack
by FGSM against an initial LS attack. 
This reveals that re-attack corrects the label despite adding
substantial perturbation  to the back of the vehicle through the
pixel greedy method.

\subsubsection{
Analysis of perturbation amounts
}
\label{ssec:analysis_perturb-amount}

This section examines the underlying reasons our proposed method can
correct AEs produced without relying on gradients, such as those from
black-box attacks, like LS and HSJA.
As discussed in
Section~\ref{ssec:theoretical_analysis}, effective
rectification of AEs requires applying a re-attack with perturbations
of appropriate direction and magnitude.
Therefore, we first conducted an analysis centered on the magnitude of
the perturbations introduced during the re-attack.

Tables~\ref{tbl:attack_norm} and \ref{tbl:re_attack_norm} show the
perturbation amounts of the initial attack for generating AEs and
re-attack for rectification, respectively.
The former represents $\| {\bm \delta} \|_2$ averaged over 1,000 AEs,
i.e., the perturbation amount required to change the label when
attacking the original sample,
and the latter represents $\| {\bm \delta}' \|_2$ averaged over 1,000
rectified AEs, i.e., the perturbation amount required to correct the
label when rectifying the AEs.
Note that all values in Table \ref{tbl:re_attack_norm} are multiplied
by $10^{-3}$.
By comparing the perturbation amounts in
the two tables, we confirmed that the latter is extremely small.

Indeed, the 
norms of AEs
generated using LS were larger than those generated via other attack
methods, as indicated in Table~\ref{tbl:attack_norm}.
The norms of AEs generated by HSJA on ImageNet were
extensive, possibly linked to HSJA occasionally failing to find
suitable AEs for certain classifiers or
inputs~\cite{chen2020hopskipjumpattack,vo2021ramboattack}.
Remarkably, the proposed method successfully estimated 
the original input labels even when dealing with AEs featuring substantial
perturbations, i.e., those substantially deviate from the original
image.

\subsubsection{%
Analysis of the appropriateness of re-attack direction
}
\label{ssec:analysis_perturb-direction}

Following Section~\ref{ssec:analysis_perturb-amount}, this section
examines why our method effectively rectifies AEs generated by
black-box attacks.
We focus on validating the appropriateness of the re-attack
perturbation's direction, specifically assessing how closely it
opposes the original perturbation ${\bm \delta}$ used to create an AE.

Table~\ref{tbl:cos_sim} presents the cosine similarity between attack
perturbation and the inverse of re-attack perturbation, which is calculated as
follows:
\begin{displaymath}
\rm{sim}_{\rm{cos}}=\frac{(-{\bm \delta},{\bm \delta'})}{||-{\bm \delta}||\cdot||{\bm \delta'}||} \;\; s.t. \;\; {\bm x_{a}}-{\bm x}={\bm \delta}, \;\; {\bm x'_{a}}-{\bm x_{a}}={\bm \delta'}
\end{displaymath}
Note that the re-attack perturbation direction was 
inverted,
and as
the similarity increased, the direction of the attack and re-attack
would be more opposite.
AEs were generated to decrease the confidence of the correct category
during the attack phase, and AEs were rectified to decrease the
confidence of the misrecognized category during the re-attack phase.
Even when employing the same white-box attack techniques for the
attack and re-attack phases, the resulting perturbations may not
necessarily be oriented in opposing directions.

Table~\ref{tbl:cos_sim} showed that AEs generated via methods that
search for perturbations in the gradient direction (FGSM, BIM, DF, and
CW) showed higher cosine similarity,
considering the high dimensionality of perturbations.
This indicates that AEs were rectified in the direction opposite to that
of the original attack, 
thus confirming our insights in Sec.~\ref{ssec:theoretical_analysis}
.

Similarities of the methods that generate pixel-wise greedy
perturbations (JSMA, LS) were lower than those of the gradient-based
methods.
This is because local perturbations suppressed visibility but did not
generate perturbations in the shortest possible distance such as in
the inverse direction of the re-attack.

Interestingly, 
HSJA, which does not utilize gradients, showed high similarity values 
on MNIST and 
CIFAR-10.
This may be attributed to its 
ability
 to search for the
optimal perturbation in a small region centered on the line segment
connecting the input and starting point.
The cosine similarity of HSJA on ImageNet is closer
to zero, attributed to the expansion of the search space and
complexity of decision boundaries owing to an increase in the number of
pixels.
Even when the similarity is near zero, meaning it is orthogonal to the
original attack perturbation direction, the proposed method can still
rectify the AE by automatically adjusting the perturbation amount, if
the AE exists in a convex adversarial region that protrudes toward the
non-adversarial region.

The use of the proposed method in this manner may suggest some
potential for it to serve as a metric for characterizing AEs,
particularly leveraging its advantage of being independent of specific
tasks or data modalities, though more investigation is needed.

\begin{figure}[t]
  \centering
    \hspace*{-2mm}\includegraphics[width=9.0cm]{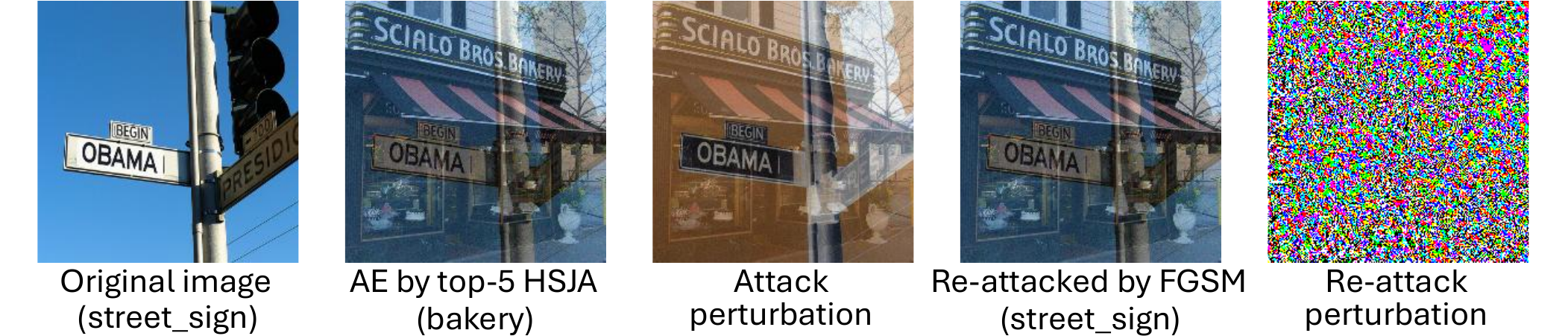}\\
    \hspace*{-2mm}\includegraphics[width=9.0cm]{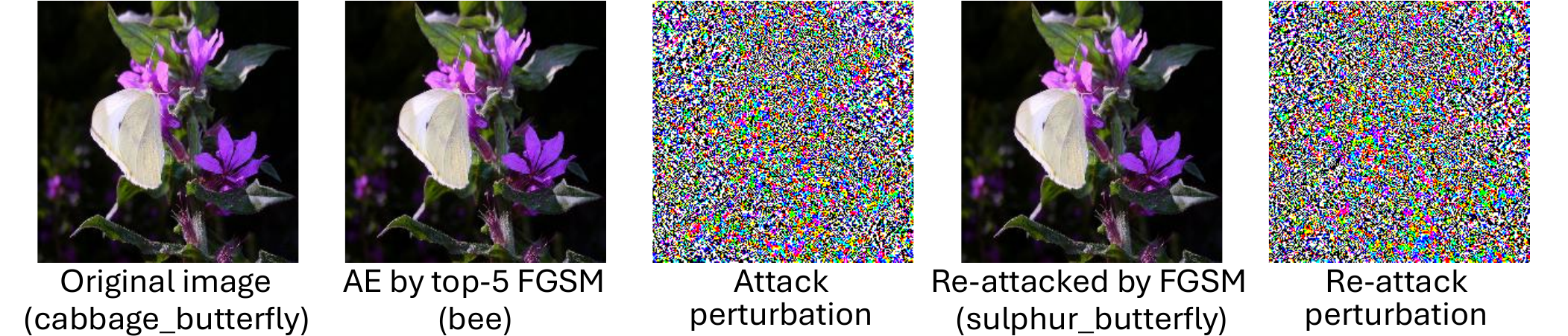}
    \caption{Example AEs generated by attack targeting Top-5 label in Experiment 1b.}
    \label{fig:exp3_exam}
\end{figure}

\begin{figure}[t]
  \centering
  \begin{tabular}{@{}p{4cm}@{~~~~~}p{4cm}@{}}
  \includegraphics[width=4cm]{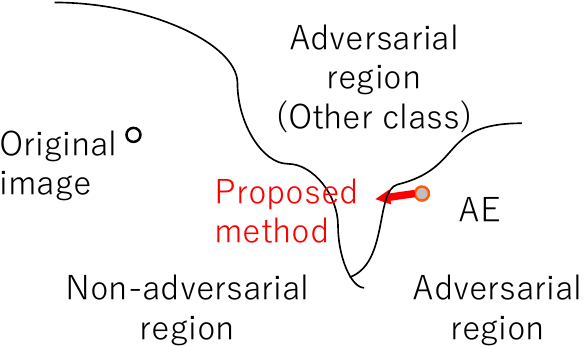} &
  \includegraphics[width=4cm]{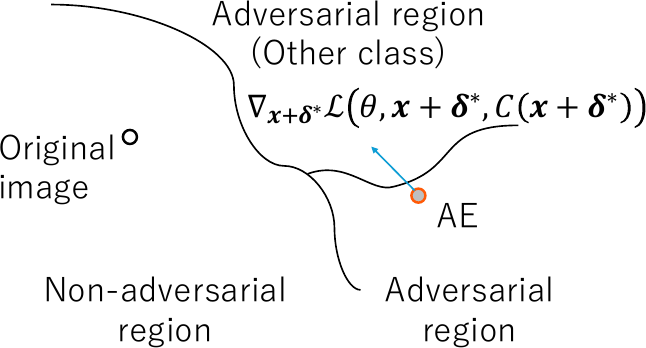} \\
  {\footnotesize \narrowminipage{
      (a) A case where misclassified and original class regions are
      not adjacent due to the presence of other class regions.
  }} &
  {\footnotesize \narrowminipage{
    (b) A case where the perturbation direction of re-attacking is not adequate. 
  }}
  \end{tabular}
  \caption{
Cases where rectification fails for AEs generated by
    targeted attacks that induce misclassification into low-ranking
    classes.}
  \label{fig:failure_cases}
~\\ ~\\
  \centering
  \begin{tabular}{@{}p{4cm}@{~~~~~}p{4cm}@{}}
  \includegraphics[width=4.8cm]{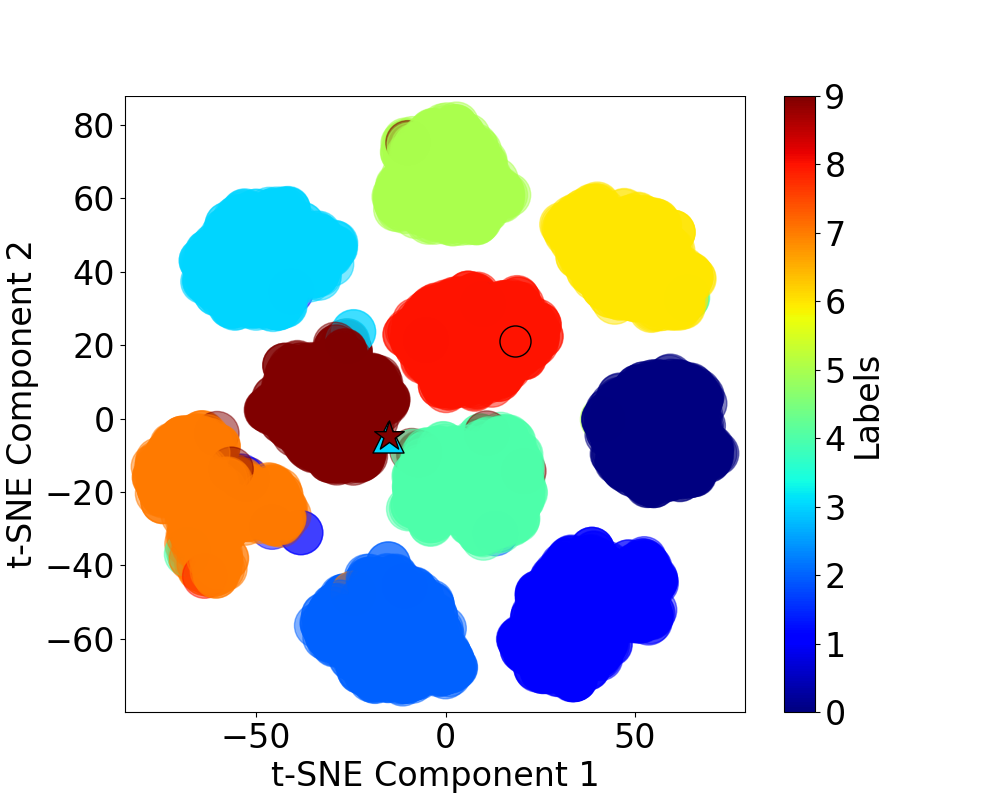} &
  \includegraphics[width=4.8cm]{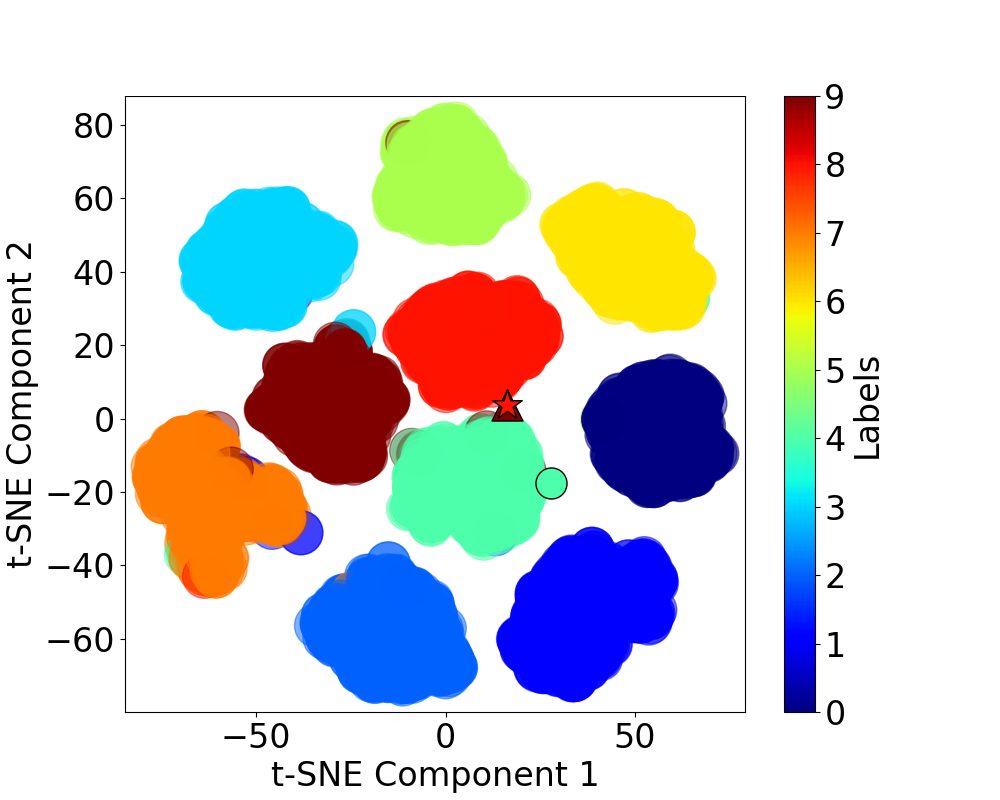} \\
  {\footnotesize \narrowminipage{
    (a) A case where misclassified and original class regions are not
      adjacent due to the presence of other class regions.
  }}
  &
  {\footnotesize \narrowminipage{
    (b) A case where the perturbation direction of re-attacking is not
      adequate.
  }}
  \end{tabular}
  \caption{ Visualization results where rectification fails for AEs
  generated by targeted attacks to low-ranking classes.
}
  \label{fig:failure_cases_tsne}
\end{figure}

\subsection{Experiment 1b: Targeted attack}

\subsubsection{%
Setup
}

In Experiment 1b, we validated the proposed method against AEs 
generated via targeted attacks.
As described in Sec.~\ref{ssec:concept},
the rectification of AEs generated by targeted attacks is expected to be
more challenging than those generated using untargeted attacks.
This difficulty arises because AEs generated by targeted attacks are
usually farther away from the original inputs.

For this experimental test, we utilized  
the Top-2 to Top-5 labels in 
the outputs of
 the classification models
as the target labels%
\footnote{
Note that Top-1 corresponds to the correct label of an input.
};
this approach is expected to increase the rectification difficulty as
the rank decreases.
The datasets, classification models, and evaluation criteria remained
consistent with Experiment 1a.
Note that our method performed untargeted re-attack in Experiment 1b,
following the same manner as in Experiment 1a, as the correct class is
unknown.

Various attack types were employed for generating AEs, including FGSM,
BIM, CW, JSMA, and HSJA.
LS was not employed in this experiment owing to its algorithm behavior
and the implementation limitation of the Foolbox framework.
Although the HSJA attack does not rely on the confidence score, it can
still perform a targeted attack by initiating from a sample belonging
to the target class and minimizing perturbations while maintaining the
classification result
\footnote{%
In this experiment, HSJA's targeted attacks were started from a
randomly selected sample classified as the target class.
}.
The attack and re-attack parameters were configured as in Experiment 1a.

\subsubsection{%
Results on rectification performance
}

Table~\ref{tbl:target_attack}
presents the
rectification success rates.
Compared with the untargeted attacks in Experiment 1a, the 
rectification success rates obtained
in Experiment 1b were lower, although the values remained high under many 
conditions.
The proposed method successfully estimated the correct labels from AEs
generated via BIM, CW, and JSMA on all three datasets and from AEs
by HSJA on MNIST and CIFAR-10.
As the target labels were changed from Top-2 to Top-5,
perturbation between the original input and AE increased,
resulting in a lower success rate.

Table~\ref{tbl:target_norm} shows the averaged perturbation amount
between the original input and AEs.
We observed that AEs generated via FGSM against Top-3 or lower target
labels on all datasets and AEs generated by HSJA on ImageNet included
large perturbations compared to those generated by other methods.
Consequently, their associated success rates were lower than those for
AEs generated under other conditions.
The FGSM targeted attack is particularly a threat to the proposed method. 
Unlike other iterative optimization methods, FGSM adds perturbations
in a straight line toward the target class, resulting in the movement
of AEs to regions in the feature space that were not rectifiable.

Figure~\ref{fig:exp3_exam} depicts two examples in which our
rectification method was applied to AEs generated by targeting the
Top-5 label.
The example in the first row corresponded to a re-attack by FGSM
following an initial attack by HSJA.
It demonstrated correct AE rectification, despite adding
large perturbation to the original input.
Meanwhile, the example in the second row illustrated a re-attack with
FGSM following an initial attack with FGSM.
In this instance, our method failed to 
correct the label, although perturbation 
of the AE
was not apparent.

\begin{table*}[t]
  \centering
  \caption{Experiment 2a: Comparison with the state-of-the-art rectification method using XAI~\cite{kao2022rectifying} and image transformation methods.}
  \label{tbl:comparison_XAI}
  {
  \begin{tabular}{@{~}c@{~}||@{~}l@{~}|@{~}l@{~}|@{~}c@{~~}c@{~~}c@{~~}c@{~}}
  \hline
  \multirow{3}{*}{Dataset} & \multicolumn{1}{@{~}c@{~}|@{~}}{\multirow{3}{*}{Approach}} &
  \multicolumn{1}{@{~}c@{~}|@{~}}{\multirow{3}{*}{Method}} & 
  \multicolumn{4}{@{~}c@{~}}{Attack method} \\ \cline{4-7}
  \multirow{1}{*}{}   & &
  \multicolumn{1}{@{~}c@{~}|@{~}}{\multirow{2}{*}{}} & 
  FGSM & BIM & BIM & CW \\
  \multirow{1}{*}{}   & &
  \multicolumn{1}{@{~}c@{~}|@{~}}{\multirow{1}{*}{}} & 
  ($L_{\infty}$) & ($L_{2}$) & ($L_{\infty}$) & ($L_{2}$) \\ 
  \hline\hline
  \multirow{8}{*}{MNIST} & \multirow{4}{*}{Input transformation} 
  & Denoising autoencoder~\cite{meng2017magnet} & 0.500 & 0.760 & 0.581 & 0.621
  \\ 
  && JPEG compression~\cite{dziugaite2016study} & 0.037 & 0.111 & 0.095 & 0.000
  \\ 
  && Full-image Gaussian blur~\cite{kao2022rectifying} & 0.389 & 0.616 & 0.459 & 0.389
  \\ 
  && Random pixel replacement~\cite{kao2022rectifying} & 0.280 & 0.320 & 0.280 & 0.260
  \\ \cline{2-7}
  & XAI-based rectification
  &Previous method (Kao et al.)~\cite{kao2022rectifying}                     & 0.889 & 0.949 & 0.905 & 0.972 \\ 
  \cline{2-7}
  & \multirow{3}{*}{Re-attack-based rectification}
  & Proposed method (FGSM)                        & \textbf{0.999} & \textbf{0.996} & \textbf{0.999} & \textbf{1.000} \\
  && Proposed method (BIM)                        & 0.998 & \textbf{0.996} & \textbf{0.999} & \textbf{1.000} \\
  && Proposed method (DF)                        & 0.993 & 0.992 & 0.998 & \textbf{1.000} \\ \hline
  \hline
  \multirow{8}{*}{CIFAR-10}
  & \multirow{4}{*}{Input transformation}
  &Denoising autoencoder~\cite{meng2017magnet} & 0.455 & 0.446 & 0.617 & 0.731
  \\ 
  && JPEG compression~\cite{dziugaite2016study} & 0.093 & 0.000 & 0.051 & 0.404
  \\ 
  && Full-image Gaussian blur~\cite{kao2022rectifying} & 0.279 & 0.271 & 0.322 & 0.277
  \\ 
  && Random pixel replacement~\cite{kao2022rectifying} & 0.190 & 0.120 & 0.180 & 0.250
  \\ \cline{2-7}
  & XAI-based rectification
  &Previous method (Kao et al.)~\cite{kao2022rectifying}                     & 0.581 & 0.616 & 0.729 & 0.936 
  \\\cline{2-7}
  & \multirow{3}{*}{Re-attack-based rectification}
  &Proposed method (FGSM)                        & 0.990 & \textbf{0.997} & \textbf{1.000} & \textbf{1.000} \\
  &&Proposed method (BIM)                       & \textbf{0.992} & \textbf{0.997} & \textbf{1.000} & \textbf{1.000} \\
  &&Proposed method (DF)                        & 0.991 & 0.995 & 0.997 & 0.998 \\ \hline
  \end{tabular}
  }
\end{table*}

\subsubsection{%
Visualization of the feature space using t-SNE
}

The difficulty in rectifying AEs misclassified into low-ranking
classes, such as Top-4 or -5, can be attributed to factors illustrated
in Fig.~\ref{fig:failure_cases}(a) and (b).
In Fig.~\ref{fig:failure_cases}(a), AEs lie in regions where the
predicted class $C({\bm x} + {\bm \delta})$ and the original class
$C({\bm x})$ are not adjacent, with other class regions in between.
Additionally, Fig.~\ref{fig:failure_cases}(b) illustrates that even if
$C({\bm x} + {\bm \delta})$ and $C({\bm x})$ are adjacent, an improper
re-attack perturbation $-\nabla_{{\bm x}+{\bm \delta^*}} L(\theta,
{{\bm x}+{\bm \delta^*}}, C({\bm x}+{\bm \delta^*})) $ can direct
perturbations away from the original class, hindering rectification.

Using t-SNE~\cite{van2008visualizing} in Experiment 1b, we visualized
the positional relationships of AEs and their rectification results in
the feature space to investigate failures in rectifying AEs
misclassified to low-rank classes.
Fig.~\ref{fig:failure_cases_tsne} illustrates the cases on the
CIFAR-10 dataset where FGSM-generated Top-5 AEs could not be
successfully rectified using FGSM.
Each class is color-coded, and all benign samples that the classifier 
was able to correctly classify are drawn as large
circles to indicate pseudo-classification regions.
Original inputs are indicated by circles, AEs by triangles, and
re-attacked AEs by stars, with a filled color indicating a class.
For example, if the circle and star have the same color, the
rectification is successful.

The example in Fig.~\ref{fig:failure_cases_tsne} demonstrates that the
generated AE (light blue class) could not transition into the original
class region (red) via re-attack, instead migrating into another class
region (brown) situated between the misclassified and original
classes, which corresponds to a case shown in
Fig.~\ref{fig:failure_cases}(a).
Similarly, Fig.~\ref{fig:failure_cases_tsne}(b) shows an example where,
despite its proximity to the original class region 
(light blue green),
the AE was mistakenly moved into a different class region (red) after
re-attack, categorizing it as a case shown in
Fig.~\ref{fig:failure_cases}(b).

\begin{table*}[t]
  \centering
  \caption{Experiment 2b: Comparison with RS\&V~\cite{wang2022detecting}, the state-of-the-art adversarial defense method for natural language processing DNNs.}
  \label{tbl:rsv_comparison}
  {
  \begin{tabular}{@{~~}c@{~~}||@{~~}l@{~~}|@{~~}c@{~~}c@{~~}c@{~~}c@{~~}c@{~~}|@{~~}c@{~~}c@{~~}}
  \hline
  \multirow{3}{*}{Dataset} &  
  \multirow{3}{*}{Defense method} & 
  \multicolumn{7}{@{~~}c@{~~}}{\multirow{1}{*}{Attack method}} \\
  \cline{3-9}
  &
  & 
  \multicolumn{5}{@{~~}c@{~~}|@{~~}}{\multirow{1}{*}{White-box}} &
  \multicolumn{2}{@{~~}c@{~~}}{\multirow{1}{*}{Black-box}} \\
  \cline{3-9} 
  & \multicolumn{1}{@{~~}c@{~~}|@{~~}}{}
  & \multicolumn{1}{@{~~}c@{~~}}{FGSM} 
  & \multicolumn{1}{@{~~}c@{~~}}{BIM} 
  & \multicolumn{1}{@{~~}c@{~~}}{DF} 
  & \multicolumn{1}{@{~~}c@{~~}}{CW} 
  & \multicolumn{1}{@{~~}c@{~~}|@{~~}}{JSMA} 
  & \multicolumn{1}{@{~~}c@{~~}}{LS} 
  & \multicolumn{1}{@{~~}c@{~~}}{HSJA} \\
  \hline\hline
  \multirow{3}{*}{MNIST} 
  &\multicolumn{1}{@{~~}l@{~~}|@{~~}}{Proposed method (FGSM)} & \textbf{0.999} & \textbf{0.999} & \textbf{0.978} & \textbf{1.000} & \textbf{0.993} & \textbf{0.938} & \textbf{1.000} \\
  \cline{2-9}
  &\multicolumn{1}{@{~~}l@{~~}|@{~~}}{RS\&V($p=0.001$)} & 0.002 & 0.078 & 0.530 & 0.169 & 0.000 & 0.000 & 0.551 \\
  &\multicolumn{1}{@{~~}l@{~~}|@{~~}}{RS\&V($p=0.01$)} & 0.060 & 0.201 & 0.532 & 0.809 & 0.005 & 0.001 & 0.963 \\
  &\multicolumn{1}{@{~~}l@{~~}|@{~~}}{RS\&V($p=0.1$)} & 0.579 & 0.649 & 0.542 & 0.998 & 0.083 & 0.006 & 0.996 \\
  &\multicolumn{1}{@{~~}l@{~~}|@{~~}}{RS\&V($p=1.0$)} & 0.918 & 0.990 & 0.576 & 0.999 & 0.298 & 0.038 & 0.999 \\
  &\multicolumn{1}{@{~~}l@{~~}|@{~~}}{RS\&V($p=10.0$)} & 0.548 & 0.770 & 0.531 & 0.775 & 0.407 & 0.206 & 0.650 \\
  \hline
  \hline
  \multirow{3}{*}{CIFAR-10} 
  &\multicolumn{1}{@{~~}l@{~~}|@{~~}}{Proposed method (FGSM)} & \textbf{0.992} & \textbf{1.000} & \textbf{1.000} & \textbf{1.000} & \textbf{0.994} & \textbf{0.911} & \textbf{1.000} \\
  \cline{2-9}
  &\multicolumn{1}{@{~~}l@{~~}|@{~~}}{RS\&V($p=0.001$)} & 0.001 & 0.139 & 0.032 & 0.038 & 0.004 & 0.001 & 0.525 \\
  &\multicolumn{1}{@{~~}l@{~~}|@{~~}}{RS\&V($p=0.01$)} & 0.006 & 0.288 & 0.078 & 0.138 & 0.008 & 0.003 & 0.552 \\
  &\multicolumn{1}{@{~~}l@{~~}|@{~~}}{RS\&V($p=0.1$)} & 0.079 & 0.787 & 0.457 & 0.788 & 0.072 & 0.016 & 0.718 \\
  &\multicolumn{1}{@{~~}l@{~~}|@{~~}}{RS\&V($p=1.0$)} & 0.570 & 0.797 & 0.696 & 0.818 & 0.589 & 0.170 & 0.683 \\
  &\multicolumn{1}{@{~~}l@{~~}|@{~~}}{RS\&V($p=10.0$)} & 0.139 & 0.152 & 0.147 & 0.149 & 0.149 & 0.040 & 0.144 \\
  \hline\hline
  \multirow{3}{*}{ImageNet}
  &\multicolumn{1}{@{~~}l@{~~}|@{~~}}{Proposed method (FGSM)} & \textbf{0.926} & \textbf{0.991} & \textbf{0.999} & \textbf{0.994} & \textbf{0.999} & \textbf{0.981} & \textbf{0.997} \\
  \cline{2-9}
  &\multicolumn{1}{@{~~}l@{~~}|@{~~}}{RS\&V($p=0.001$)} & 0.003 & 0.113 & 0.076 & 0.008 & 0.013 & 0.007 & 0.633 \\
  &\multicolumn{1}{@{~~}l@{~~}|@{~~}}{RS\&V($p=0.01$)} & 0.004 & 0.324 & 0.180 & 0.023 & 0.015 & 0.009 & 0.688 \\
  &\multicolumn{1}{@{~~}l@{~~}|@{~~}}{RS\&V($p=0.1$)} & 0.040 & 0.974 & 0.722 & 0.349 & 0.100 & 0.039 & 0.629 \\
  &\multicolumn{1}{@{~~}l@{~~}|@{~~}}{RS\&V($p=1.0$)} & 0.761 & 0.988 & 0.973 & 0.886 & 0.701 & 0.266 & 0.772 \\
  &\multicolumn{1}{@{~~}l@{~~}|@{~~}}{RS\&V($p=10.0$)} & 0.909 & 0.919 & 0.921 & 0.921 & 0.891 & 0.421 & 0.823 \\
  \hline
  \end{tabular}
  }
\end{table*}

\begin{table}[t]
  \centering
  \caption{Experiment 3: Detection accuracy of $\rm A^2D$ using Z-score.}
  \label{tbl:reproduce_a2d_zscore}
  {
  \begin{tabular}{@{~}c@{~}||@{~}c@{~}|@{~}c@{~}|@{~}c@{~}|@{~}c@{~}|@{~}c@{~}|@{~}c@{~}|@{~}c@{~}}
  \hline
  \multicolumn{1}{@{~}c@{~}||@{~}}{\multirow{2}{*}{Dataset}} & 
  \multicolumn{1}{@{~}c@{~}|@{~}}{\multirow{2}{*}{Detector}} & 
  \multicolumn{6}{@{~}c@{~}}{\multirow{1}{*}{Detection accuracy}} \\
  \cline{3-8}
  \multicolumn{1}{@{~}c@{~}||@{~}}{\multirow{1}{*}{}} & 
  \multicolumn{1}{@{~}c@{~}|@{~}}{\multirow{1}{*}{}} & 
  \multicolumn{1}{@{~}c@{~}|@{~}}{\multirow{1}{*}{FGSM}} &
  \multicolumn{1}{@{~}c@{~}|@{~}}{\multirow{1}{*}{BIM}} &
  \multicolumn{1}{@{~}c@{~}|@{~}}{\multirow{1}{*}{JSMA}} &
  \multicolumn{1}{@{~}c@{~}|@{~}}{\multirow{1}{*}{CW}} &
  \multicolumn{1}{@{~}c@{~}|@{~}}{\multirow{1}{*}{$\rm{Avg}_{a}$}} &
  \multicolumn{1}{@{~}c@{~}}{\multirow{1}{*}{$\rm{bng}$}} \\
  \hline
  \hline
  \multicolumn{1}{@{~}c@{~}||@{~}}{\multirow{4}{*}{MNIST}} & 
  \multicolumn{1}{@{~}c@{~}|@{~}}{\multirow{1}{*}{BIM}} & 
  \multicolumn{1}{@{~}c@{~}|@{~}}{\multirow{1}{*}{1.000}} & 
  \multicolumn{1}{@{~}c@{~}|@{~}}{\multirow{1}{*}{1.000}} & 
  \multicolumn{1}{@{~}c@{~}|@{~}}{\multirow{1}{*}{0.999}} & 
  \multicolumn{1}{@{~}c@{~}|@{~}}{\multirow{1}{*}{1.000}} & 
  \multicolumn{1}{@{~}c@{~}|@{~}}{\multirow{1}{*}{0.999}} & 
  \multicolumn{1}{@{~}c@{~}}{\multirow{1}{*}{0.862}} \\ 
  \cline{2-8}
  \multicolumn{1}{@{~}c@{~}||@{~}}{\multirow{4}{*}{}} & 
  \multicolumn{1}{@{~}c@{~}|@{~}}{\multirow{1}{*}{BIM($L_2$)}} &  
  \multicolumn{1}{@{~}c@{~}|@{~}}{\multirow{1}{*}{1.000}} & 
  \multicolumn{1}{@{~}c@{~}|@{~}}{\multirow{1}{*}{1.000}} & 
  \multicolumn{1}{@{~}c@{~}|@{~}}{\multirow{1}{*}{0.998}} & 
  \multicolumn{1}{@{~}c@{~}|@{~}}{\multirow{1}{*}{1.000}} & 
  \multicolumn{1}{@{~}c@{~}|@{~}}{\multirow{1}{*}{0.999}} & 
  \multicolumn{1}{@{~}c@{~}}{\multirow{1}{*}{0.827}} \\
  \cline{2-8}
  \multicolumn{1}{@{~}c@{~}||@{~}}{\multirow{4}{*}{}} & 
  \multicolumn{1}{@{~}c@{~}|@{~}}{\multirow{1}{*}{JSMA}} & 
  \multicolumn{1}{@{~}c@{~}|@{~}}{\multirow{1}{*}{1.000}} & 
  \multicolumn{1}{@{~}c@{~}|@{~}}{\multirow{1}{*}{0.997}} & 
  \multicolumn{1}{@{~}c@{~}|@{~}}{\multirow{1}{*}{0.999}} & 
  \multicolumn{1}{@{~}c@{~}|@{~}}{\multirow{1}{*}{1.000}} & 
  \multicolumn{1}{@{~}c@{~}|@{~}}{\multirow{1}{*}{0.999}} & 
  \multicolumn{1}{@{~}c@{~}}{\multirow{1}{*}{0.881}} \\
  \cline{2-8}
  \multicolumn{1}{@{~}c@{~}||@{~}}{\multirow{4}{*}{}} & 
  \multicolumn{1}{@{~}c@{~}|@{~}}{\multirow{1}{*}{DBA}} &  
  \multicolumn{1}{@{~}c@{~}|@{~}}{\multirow{1}{*}{0.940}} & 
  \multicolumn{1}{@{~}c@{~}|@{~}}{\multirow{1}{*}{0.969}} & 
  \multicolumn{1}{@{~}c@{~}|@{~}}{\multirow{1}{*}{0.926}} & 
  \multicolumn{1}{@{~}c@{~}|@{~}}{\multirow{1}{*}{0.988}} & 
  \multicolumn{1}{@{~}c@{~}|@{~}}{\multirow{1}{*}{0.955}} & 
  \multicolumn{1}{@{~}c@{~}}{\multirow{1}{*}{0.919}} \\
  \hline
  \hline
  \multicolumn{1}{@{~}c@{~}||@{~}}{\multirow{4}{*}{CIFAR-10}} & 
  \multicolumn{1}{@{~}c@{~}|@{~}}{\multirow{1}{*}{BIM}} & 
  \multicolumn{1}{@{~}c@{~}|@{~}}{\multirow{1}{*}{0.812}} & 
  \multicolumn{1}{@{~}c@{~}|@{~}}{\multirow{1}{*}{0.997}} & 
  \multicolumn{1}{@{~}c@{~}|@{~}}{\multirow{1}{*}{0.956}} & 
  \multicolumn{1}{@{~}c@{~}|@{~}}{\multirow{1}{*}{0.996}} & 
  \multicolumn{1}{@{~}c@{~}|@{~}}{\multirow{1}{*}{0.940}} & 
  \multicolumn{1}{@{~}c@{~}}{\multirow{1}{*}{0.866}} \\
  \cline{2-8}
  \multicolumn{1}{@{~}c@{~}||@{~}}{\multirow{4}{*}{}} & 
  \multicolumn{1}{@{~}c@{~}|@{~}}{\multirow{1}{*}{BIM($L_2$)}} & 
  \multicolumn{1}{@{~}c@{~}|@{~}}{\multirow{1}{*}{0.834}} & 
  \multicolumn{1}{@{~}c@{~}|@{~}}{\multirow{1}{*}{0.997}} & 
  \multicolumn{1}{@{~}c@{~}|@{~}}{\multirow{1}{*}{0.963}} & 
  \multicolumn{1}{@{~}c@{~}|@{~}}{\multirow{1}{*}{0.996}} & 
  \multicolumn{1}{@{~}c@{~}|@{~}}{\multirow{1}{*}{0.947}} & 
  \multicolumn{1}{@{~}c@{~}}{\multirow{1}{*}{0.856}} \\ 
  \cline{2-8}
  \multicolumn{1}{@{~}c@{~}||@{~}}{\multirow{4}{*}{}} & 
  \multicolumn{1}{@{~}c@{~}|@{~}}{\multirow{1}{*}{JSMA}} & 
  \multicolumn{1}{@{~}c@{~}|@{~}}{\multirow{1}{*}{0.843}} & 
  \multicolumn{1}{@{~}c@{~}|@{~}}{\multirow{1}{*}{0.999}} & 
  \multicolumn{1}{@{~}c@{~}|@{~}}{\multirow{1}{*}{0.971}} & 
  \multicolumn{1}{@{~}c@{~}|@{~}}{\multirow{1}{*}{1.000}} & 
  \multicolumn{1}{@{~}c@{~}|@{~}}{\multirow{1}{*}{0.953}} & 
  \multicolumn{1}{@{~}c@{~}}{\multirow{1}{*}{0.857}} \\
  \cline{2-8}
  \multicolumn{1}{@{~}c@{~}||@{~}}{\multirow{4}{*}{}} & 
  \multicolumn{1}{@{~}c@{~}|@{~}}{\multirow{1}{*}{DBA}} & 
  \multicolumn{1}{@{~}c@{~}|@{~}}{\multirow{1}{*}{0.524}} & 
  \multicolumn{1}{@{~}c@{~}|@{~}}{\multirow{1}{*}{0.961}} & 
  \multicolumn{1}{@{~}c@{~}|@{~}}{\multirow{1}{*}{0.651}} & 
  \multicolumn{1}{@{~}c@{~}|@{~}}{\multirow{1}{*}{0.967}} & 
  \multicolumn{1}{@{~}c@{~}|@{~}}{\multirow{1}{*}{0.775}} & 
  \multicolumn{1}{@{~}c@{~}}{\multirow{1}{*}{0.964}} \\
  \hline
  \hline
  \multicolumn{1}{@{~}c@{~}||@{~}}{\multirow{4}{*}{ImageNet}} & 
  \multicolumn{1}{@{~}c@{~}|@{~}}{\multirow{1}{*}{BIM}} & 
  \multicolumn{1}{@{~}c@{~}|@{~}}{\multirow{1}{*}{1.000}} & 
  \multicolumn{1}{@{~}c@{~}|@{~}}{\multirow{1}{*}{0.999}} & 
  \multicolumn{1}{@{~}c@{~}|@{~}}{\multirow{1}{*}{1.000}} & 
  \multicolumn{1}{@{~}c@{~}|@{~}}{\multirow{1}{*}{0.975}} & 
  \multicolumn{1}{@{~}c@{~}|@{~}}{\multirow{1}{*}{0.993}} & 
  \multicolumn{1}{@{~}c@{~}}{\multirow{1}{*}{0.876}} \\
  \cline{2-8}
  \multicolumn{1}{@{~}c@{~}||@{~}}{\multirow{4}{*}{}} & 
  \multicolumn{1}{@{~}c@{~}|@{~}}{\multirow{1}{*}{BIM($L_2$)}} & 
  \multicolumn{1}{@{~}c@{~}|@{~}}{\multirow{1}{*}{0.999}} & 
  \multicolumn{1}{@{~}c@{~}|@{~}}{\multirow{1}{*}{0.998}} & 
  \multicolumn{1}{@{~}c@{~}|@{~}}{\multirow{1}{*}{1.000}} & 
  \multicolumn{1}{@{~}c@{~}|@{~}}{\multirow{1}{*}{0.968}} & 
  \multicolumn{1}{@{~}c@{~}|@{~}}{\multirow{1}{*}{0.991}} & 
  \multicolumn{1}{@{~}c@{~}}{\multirow{1}{*}{0.882}} \\ 
  \cline{2-8}
  \multicolumn{1}{@{~}c@{~}||@{~}}{\multirow{4}{*}{}} & 
  \multicolumn{1}{@{~}c@{~}|@{~}}{\multirow{1}{*}{JSMA}} &  
  \multicolumn{1}{@{~}c@{~}|@{~}}{\multirow{1}{*}{0.986}} & 
  \multicolumn{1}{@{~}c@{~}|@{~}}{\multirow{1}{*}{0.996}} & 
  \multicolumn{1}{@{~}c@{~}|@{~}}{\multirow{1}{*}{1.000}} & 
  \multicolumn{1}{@{~}c@{~}|@{~}}{\multirow{1}{*}{0.979}} & 
  \multicolumn{1}{@{~}c@{~}|@{~}}{\multirow{1}{*}{0.990}} & 
  \multicolumn{1}{@{~}c@{~}}{\multirow{1}{*}{0.865}} \\ 
  \cline{2-8}
  \multicolumn{1}{@{~}c@{~}||@{~}}{\multirow{4}{*}{}} & 
  \multicolumn{1}{@{~}c@{~}|@{~}}{\multirow{1}{*}{DBA}} & 
  \multicolumn{1}{@{~}c@{~}|@{~}}{\multirow{1}{*}{0.940}} & 
  \multicolumn{1}{@{~}c@{~}|@{~}}{\multirow{1}{*}{0.997}} & 
  \multicolumn{1}{@{~}c@{~}|@{~}}{\multirow{1}{*}{0.970}} & 
  \multicolumn{1}{@{~}c@{~}|@{~}}{\multirow{1}{*}{0.965}} & 
  \multicolumn{1}{@{~}c@{~}|@{~}}{\multirow{1}{*}{0.968}} & 
  \multicolumn{1}{@{~}c@{~}}{\multirow{1}{*}{0.886}} \\ 
  \hline
  \end{tabular}
  }
~\\ ~\\
  \centering
  \caption{Experiment 3: classifiers' accuracy with the proposed method and $\rm A^2D$ using Z-score.}
  \label{tbl:a2d_exp}
  {
  \begin{tabular}{@{~}c@{~}||@{~}c@{~}|@{~}c@{~}|@{~}c@{~}c@{~}c@{~}c@{~}|@{~}c@{~}|@{~}c@{~}}
  \hline
  \multirow{2}{*}{Dataset} &
  \multirow{2}{*}{Detector} & 
  Re-attack &
  \multicolumn{4}{@{~}c@{~}|@{~}}{Attack method} &
  \multirow{2}{*}{$\rm{Avg}_{a}$} &
  \multirow{2}{*}{$\rm{bng}$} \\ \cline{4-7}
  && \multicolumn{1}{@{~}c@{~}|@{~}}{method} & 
  \multicolumn{1}{@{~}c@{~}}{\multirow{1}{*}{FGSM}} &
  \multicolumn{1}{@{~}c@{~}}{\multirow{1}{*}{BIM}} &
  \multicolumn{1}{@{~}c@{~}}{\multirow{1}{*}{JSMA}} &
  \multicolumn{1}{@{~}c@{~}|@{~}}{\multirow{1}{*}{CW}} &&\\
  \hline\hline
  \multirow{12}{*}{MNIST}
  &\multicolumn{1}{@{~}c@{~}|@{~}}{\multirow{3}{*}{BIM}} & \multicolumn{1}{@{~}c@{~}|@{~}}{FGSM} & 
  \textbf{0.975} & 0.983 & \textbf{0.954} & \textbf{1.000} & \textbf{0.978} & 0.862 \\
  &\multicolumn{1}{@{~}l@{~}|@{~}} {} & \multicolumn{1}{@{~}c@{~}|@{~}}{BIM} & 
  \textbf{0.975} & \textbf{0.984} & 0.951 & \textbf{1.000} & 0.976 & 0.862 \\
  &\multicolumn{1}{@{~}l@{~}|@{~}} {} & \multicolumn{1}{@{~}c@{~}|@{~}}{DF} & 
  0.899 & 0.958 & 0.915 & 0.999 & 0.943 & 0.862 \\
  \cline{2-9}
  &\multicolumn{1}{@{~}c@{~}|@{~}}{\multirow{2}{*}{BIM}} & \multicolumn{1}{@{~}c@{~}|@{~}}{FGSM} & 
  \textbf{0.975} & 0.983 & 0.953 & \textbf{1.000} & \textbf{0.978} & 0.827 \\
  &\multicolumn{1}{@{~}c@{~}|@{~}}{\multirow{2}{*}{($L_2$)}} & \multicolumn{1}{@{~}c@{~}|@{~}}{BIM} & 
  \textbf{0.975} & \textbf{0.984} & 0.951 & \textbf{1.000} & \textbf{0.978} & 0.827 \\
  &\multicolumn{1}{@{~}l@{~}|@{~}} {} & \multicolumn{1}{@{~}c@{~}|@{~}}{DF} & 
  0.887 & 0.958 & 0.902 & 0.997 & 0.936 & 0.827 \\
  \cline{2-9}
  &\multicolumn{1}{@{~}c@{~}|@{~}}{\multirow{3}{*}{JSMA}} & \multicolumn{1}{@{~}c@{~}|@{~}}{FGSM} & 
  \textbf{0.975} & 0.981 & \textbf{0.954} & \textbf{1.000} & \textbf{0.978} & 0.881 \\
  &\multicolumn{1}{@{~}l@{~}|@{~}} {} & \multicolumn{1}{@{~}c@{~}|@{~}}{BIM} & 
  \textbf{0.975} & 0.982 & 0.951 & \textbf{1.000} & 0.977 & 0.881 \\
  &\multicolumn{1}{@{~}l@{~}|@{~}} {} & \multicolumn{1}{@{~}c@{~}|@{~}}{DF} & 
  0.910 & 0.953 & 0.910 & 0.998 & 0.943 & 0.881 \\
  \cline{2-9}
  &\multicolumn{1}{@{~}c@{~}|@{~}}{\multirow{3}{*}{DBA}} & \multicolumn{1}{@{~}c@{~}|@{~}}{FGSM} & 
  0.917 & 0.953 & 0.882 & 0.988 & 0.935 & \textbf{0.919} \\
  &\multicolumn{1}{@{~}l@{~}|@{~}} {} & \multicolumn{1}{@{~}c@{~}|@{~}}{BIM} & 
  0.917 & 0.954 & 0.879 & 0.988 & 0.935 & \textbf{0.919} \\
  &\multicolumn{1}{@{~}l@{~}|@{~}} {} & \multicolumn{1}{@{~}c@{~}|@{~}}{DF} & 
  0.845 & 0.930 & 0.847 & 0.986 & 0.902 & \textbf{0.919} \\
  \hline \hline
  \multirow{11}{*}{CIFAR-} 
  &\multicolumn{1}{@{~}c@{~}|@{~}}{\multirow{3}{*}{BIM}} 
   & FGSM & 0.632 & 0.986 & 0.936 & \textbf{0.995} & 0.887 & 0.866 \\
  \multirow{11}{*}{10} 
  && BIM & 0.631 & 0.990 & 0.935 & \textbf{0.995} & 0.888 & 0.866 \\
  && DF  & 0.639 & 0.989 & 0.939 & 0.991 & 0.890 & 0.866 \\
  \cline{2-9}
  &\multicolumn{1}{@{~}c@{~}|@{~}}{\multirow{2}{*}{BIM}} & \multicolumn{1}{@{~}c@{~}|@{~}}{FGSM} & 
  0.649 & 0.986 & 0.942 & \textbf{0.995} & 0.893 & 0.856 \\
  &\multicolumn{1}{@{~}c@{~}|@{~}}{\multirow{2}{*}{($L_2$)}} & \multicolumn{1}{@{~}c@{~}|@{~}}{BIM} & 
  0.649 & 0.990 & 0.941 & \textbf{0.995} & 0.894 & 0.856 \\
  &\multicolumn{1}{@{~}l@{~}|@{~}} {} & \multicolumn{1}{@{~}c@{~}|@{~}}{DF} & 
  0.655 & 0.989 & 0.945 & 0.991 & 0.895 & 0.856 \\
  \cline{2-9}
  &\multicolumn{1}{@{~}c@{~}|@{~}}{\multirow{3}{*}{JSMA}} & \multicolumn{1}{@{~}c@{~}|@{~}}{FGSM} & 
  0.666 & 0.986 & 0.948 & \textbf{0.995} & 0.899 & 0.857 \\
  &\multicolumn{1}{@{~}l@{~}|@{~}} {} & \multicolumn{1}{@{~}c@{~}|@{~}}{BIM} & 
  0.663 & 0.990 & 0.948 & \textbf{0.995} & 0.899 & 0.857 \\
  &\multicolumn{1}{@{~}l@{~}|@{~}} {} & \multicolumn{1}{@{~}c@{~}|@{~}}{DF} & 
  \textbf{0.676} & \textbf{0.991} & \textbf{0.953} & \textbf{0.995} & \textbf{0.904} & 0.857 \\
  \cline{2-9}
  &\multicolumn{1}{@{~}c@{~}|@{~}}{\multirow{3}{*}{DBA}} & \multicolumn{1}{@{~}c@{~}|@{~}}{FGSM} & 
  0.416 & 0.948 & 0.639 & 0.962 & 0.741 & \textbf{0.964} \\
  &\multicolumn{1}{@{~}l@{~}|@{~}} {} & \multicolumn{1}{@{~}c@{~}|@{~}}{BIM} & 
  0.418 & 0.952 & 0.639 & 0.962 & 0.743 & \textbf{0.964} \\
  &\multicolumn{1}{@{~}l@{~}|@{~}} {} & \multicolumn{1}{@{~}c@{~}|@{~}}{DF} & 
  0.427 & 0.953 & 0.639 & 0.962 & 0.745 & \textbf{0.964} \\
  \hline\hline
  \multirow{11}{*}{Image-} 
  &\multicolumn{1}{@{~}c@{~}|@{~}}{\multirow{3}{*}{BIM}} & \multicolumn{1}{@{~}c@{~}|@{~}}{FGSM} & 
  0.954 & 0.994 & \textbf{1.000} & 0.973 & 0.980 & 0.876 \\
  \multirow{11}{*}{Net}
  &\multicolumn{1}{@{~}l@{~}|@{~}} {} & \multicolumn{1}{@{~}c@{~}|@{~}}{BIM} & 
  \textbf{0.956} & \textbf{0.998} & \textbf{1.000} & 0.974 & \textbf{0.982} & 0.876 \\
  &\multicolumn{1}{@{~}l@{~}|@{~}} {} & \multicolumn{1}{@{~}c@{~}|@{~}}{DF} & 
  0.954 & 0.993 & \textbf{1.000} & 0.971 & 0.980 & 0.876 \\
  \cline{2-9} 
  &\multicolumn{1}{@{~}c@{~}|@{~}}{\multirow{2}{*}{BIM}} & \multicolumn{1}{@{~}c@{~}|@{~}}{FGSM} & 
  0.953 & 0.993 & \textbf{1.000} & 0.966 & 0.978 & 0.882 \\
  &\multicolumn{1}{@{~}c@{~}|@{~}}{\multirow{2}{*}{($L_2$)}} & \multicolumn{1}{@{~}c@{~}|@{~}}{BIM} & 
  0.955 & 0.997 & \textbf{1.000} & 0.967 & 0.980 & 0.882 \\
  &\multicolumn{1}{@{~}l@{~}|@{~}} {} & \multicolumn{1}{@{~}c@{~}|@{~}}{DF} & 
  0.952 & 0.992 & 0.999 & 0.965 & 0.977 & 0.882 \\
  \cline{2-9}
  &\multicolumn{1}{@{~}c@{~}|@{~}}{\multirow{3}{*}{JSMA}} & \multicolumn{1}{@{~}c@{~}|@{~}}{FGSM} & 
  0.940 & 0.991 & \textbf{1.000} & 0.977 & 0.977 & 0.865 \\
  &\multicolumn{1}{@{~}l@{~}|@{~}} {} & \multicolumn{1}{@{~}c@{~}|@{~}}{BIM} & 
  0.942 & 0.995 & \textbf{1.000} & \textbf{0.978} & 0.979 & 0.865 \\
  &\multicolumn{1}{@{~}l@{~}|@{~}} {} & \multicolumn{1}{@{~}c@{~}|@{~}}{DF} & 
  0.940 & 0.991 & 0.999 & 0.974 & 0.976 & 0.865 \\
  \cline{2-9}
  &\multicolumn{1}{@{~}c@{~}|@{~}}{\multirow{3}{*}{DBA}} & \multicolumn{1}{@{~}c@{~}|@{~}}{FGSM} & 
  0.896 & 0.992 & 0.970 & 0.963 & 0.955 & \textbf{0.886} \\
  &\multicolumn{1}{@{~}l@{~}|@{~}} {} & \multicolumn{1}{@{~}c@{~}|@{~}}{BIM} & 
  0.898 & 0.996 & 0.970 & 0.962 & 0.957 & \textbf{0.886} \\
  &\multicolumn{1}{@{~}l@{~}|@{~}} {} & \multicolumn{1}{@{~}c@{~}|@{~}}{DF} & 
  0.895 & 0.990 & 0.969 & 0.959 & 0.953 & \textbf{0.886} \\
  \hline
  \end{tabular}
  }
\end{table}

\subsection{Experiment 2a: Comparison with the state-of-the-art rectification method using XAI}

\jtextd{
Experiment 2では，提案手法の性能を検証するために，先行研究との比較を行った．
まず，提案手法と同様に敵対的事例の矯正に焦点を置いた手法との比較を行い
（Experiment 2a），次に，最新の自然言語処理用DNNを対象とした敵対的防御
手法の一つであるRS\&V
との比較を行った（Experiment 2b）．
}

We
compared our method with the one utilizing XAI,
which is previously reported by Kao et al~\cite{kao2022rectifying},
representing a state-of-the-art approach for rectifying detected AEs,
in addition to four image transformation methods: denoising
autoencoder~\cite{meng2017magnet}, JPEG
compression~\cite{dziugaite2016study}, full-image Gaussian
blur~\cite{kao2022rectifying},  and random pixel
replacement~\cite{kao2022rectifying}.
A denoising autoencoder adopts the reformer module
of a defense method called MagNet proposed by Meng et al.
JPEG compression, proposed by Dziugaite et al. is a defense method
that removes high-frequency components in images.
Full-image Gaussian blur is a simple image blur filter that calculates
each pixel transformation in an image using a normal
distribution.
Random pixel replacement replaces some randomly selected 
pixels of AEs 
with
 black.

For this experimental evaluation, we utilized the FGSM, BIM ($L_{2},
L_{\infty}$), and CW methods as attack types for AE generation on the
MNIST and CIFAR-10 datasets under an untargeted attack scenario.
The attack parameters for AE generation
were
configured according to a previous work~\cite{kao2022rectifying}.
The classification model for MNIST and CIFAR-10 and evaluation
criterion remained identical with Experiment 1a.

Table~\ref{tbl:comparison_XAI} lists the results of Experiment 2,
where
we referenced the best results for the method using XAI that align
with the conditions outlined in Ref.~\cite{kao2022rectifying}.
The results of the four input transformation methods were similarly
sourced from Ref.~\cite{kao2022rectifying}%
\footnote{
We attempted to align our experimental conditions as closely as
possible to theirs, however, due to undisclosed details such as
samples and classification model weights, replicating
the exact conditions was not feasible.
Differences in samples and model training details, despite using the
same dataset and DNN models, should be considered.
}.

Compared with the four input transformation methods, our proposed
method exhibited superior rectification performance that overshadows
any minor variations in the experimental setup.
To compare our method with Kao et al.'s method, we focused on the
relative success rates across different attack methods
because differences in the test samples hindered a strict comparison.
The method proposed by Kao et al. tended to decrease the rectification
success rates of AEs generated via FGSM and BIM compared with those of
AEs generated by CW on CIFAR-10 due to erroneous interpretation,
particularly evident on CIFAR-10.
This is possibly because AEs generated using BIM and FGSM contain greater
perturbations and can be further from the decision boundary than
those generated by CW, as shown in Table~\ref{tbl:attack_norm}.
Conversely, the proposed method exhibited high and consistent
rectification performance for all attacks, with no significant change
in the success rates depending on the dataset or combination of attack and
re-attack methods.

\subsection{%
Experiment 2b: Comparison with 
the state-of-the-art defense method
 in natural language processing}
\label{ssec:vs_rsv}

Next, we benchmarked the proposed method against RS\&V~\cite{wang2022detecting}, one of the defense methods designed to protect DNNs in natural language processing against AEs.
RS\&V is an inference-time defense method that, similar to our method,
can rectify AEs to their original input classes by re-attacking the
input.
It generates $k$ similar sentences by replacing words in a textual AE
with synonyms, and performs AE detection and rectification based on
the percentage of agreement between their classification results.
Because RS\&V's perturbation method is tailored for the language
modality, we modified the re-attacking in RS\&V to add random noise
to all pixels of an input
image.
Although other methods discussed in Section \ref{ssec:recent_methods}
exist, we selected RS\&V as the comparison target due to the lack of
requirement for pre-training and its applicability under conditions
similar to our proposed method.

The modified RS\&V used in Experiment 2b generated $k=25$ different
derivative images with random noise whose size was fixed in the $L_2$
norm $p$.
The choice of $k=25$ was based on the empirical findings as the
optimal number of derivation samples in RS\&V
~\cite{wang2022detecting}.
This method rectified AEs by determining the majority of the
classification results among the $k$ derived images.

The experimental setup for the proposed method
was the same as in Sec.~\ref{ssec:untargeted}, 
and the proposed method employed FGSM for re-attacking.
The RS\&V parameter $p$ varied in the range of $0.0001$ to $10.0$.

Table~\ref{tbl:rsv_comparison} shows the comparison results
with RS\&V.
The results show that RS\&V could rectify more than 90\% of AEs
generated by white-box attacks, including BIM and CW in particular,
under some conditions, although the rectification success rate of AEs
by LS was extremely low, 21\% for MNIST, 17\% for CIFAR-10, and 42\%
for ImageNet.
In contrast,
our proposed method outperformed RS\&V under all conditions 
and rectified more than 90\% of AEs by LS in all datasets.

Note that, as demonstrated in Appendix \ref{ssec:robustness}, our
proposed method requires almost no adjustment of control parameters.
Conversely, RS\&V requires the proper adjustment of $p$ in accordance
with the problem.
A smaller value of parameter $p$ hindered RS\&V's rectification of AEs
because the sampling regions were far from the region of $C({\bm x})$,
while a larger value of $p$ hindered rectification due to 
sampling of many regions from classes other than $C({\bm x})$.

Because RS\&V added random noise, the probability of successful
rectification was expected to be 50\% if an AE was located near a
decision boundary
that could be approximated by a hyperplane.
Thus, the success rectification rate by RS\&V was expected to be low;
however, it sometimes performed well against white-box attacks depending
on the parameter $p$, dataset, and attack method, which is contrary to
our intuition.
This means that many points on the hypersphere of radius $p$ centered
at an AE lie within the region of the original class $C({\bm x})$, 
indicating the AE is surrounded by a convex decision boundary that protrudes
toward the region of $C({\bm x})$.

\subsection{Experiment 3: Synergy with detector}
\label{ssec:exp4}

In Experiment 3, we evaluated the performance of our method when
integrated as a post-processing step in a conventional AE detector,
specifically $\rm A^2D$~\cite{zhao2021attack}.
Because our method is designed to be combined with a detector, if the
detector mistakenly fails to identify an AE and classifies it as a
benign sample, our method inadvertently performs a re-attack on
the benign sample, thereby generating a new AE.
This is a limitation inherent in the design of the proposed
method.
Therefore, Experiment 3 focuses on demonstrating the overall
performance when our method operates in conjunction with a detector.

In this scenario, the proposed method 
rectifies AEs detected by $\rm A^2D$.
We adopted the experimental setup outlined in
Ref.~\cite{zhao2021attack}, wherein FGSM, BIM, JSMA, and CW were
utilized as attack methods for generating AEs.
We implemented classification models for MNIST and
CIFAR-10 based on publicly available code and a  model based on
ResNet-101~\cite{he2016deep} for the ImageNet dataset.
These classifiers were trained according to the experimental setup 
in Ref.~\cite{zhao2021attack}, and were different from those utilized
in Experiments 1 and 2.
Furthermore, we configured the $\rm A^2D$ detector to utilize BIM, BIM
($L_2$), JSMA, and DBA for attacking inputs and employed Z-score and
k-NN to evaluate the robustness of inputs.

Table~\ref{tbl:reproduce_a2d_zscore} and Table~\ref{tbl:a2d_exp} show
the detection accuracies of $\rm A^2D$ with Z-score and the
classification accuracies of the classifiers equipped with $\rm A^2D$
and our proposed method, respectively.
$\rm{Avg}_{a}$ and $\rm{bng}$ denote
averaged classification accuracies for AEs generated by the four attack
methods, and accuracies for benign samples.

The comparison of the attack methods in $\rm A^2D$ revealed that the
use of DBA yielded the highest accuracy for benign samples, whereas
the accuracies against AEs were comparatively low.
The other three methods (BIM, BIM ($L_2$), and JSMA) demonstrated 
high defense performance against AEs under various conditions.
For benign samples, the classification accuracy of our method
matched the rate at which $\rm A^2D$ correctly classified them.
Note that the classification accuracy in Table~\ref{tbl:a2d_exp}
could not surpass the detection accuracy in
Table~\ref{tbl:reproduce_a2d_zscore} because the proposed method only
rectified the AEs detected by $\rm A^2D$.
For instance, when using DBA as the detector and DF as the re-attack
method, the $\rm{Avg}_{a}$ on CIFAR-10 was 0.745, which may appear lower compared to other conditions.
However, this can be attributed to the $\rm A^2D$ detection accuracy of 0.775,
as shown in Table~\ref{tbl:reproduce_a2d_zscore}.
That is, 96\% of the input AEs are successfully rectified in this
condition.
These findings indicate that the proposed method can rectify AEs
without degrading the detector performance.
When our method was combined with $\rm A^2D$ using k-NN, the same
tendency as the combination of $\rm A^2D$ with Z-score and our method
was observed.

\subsection{Experiment 4: Application to speech recognition}
\label{ssec:exp5}

To demonstrate the applicability of our proposed method beyond image
modalities, Experiment 4 was conducted to rectify AEs on a DNN
implemented for speech recognition.
BC-ResNet-8~\cite{kim2023broadcastedresiduallearningefficient}, a
convolutional neural network model known for its high accuracy on 
standard audio classification datasets, served as the victim model.
The Google Speech Commands
dataset~\cite{warden2018speechcommandsdatasetlimitedvocabulary}
containing 10 voice command classes was utilized, selecting 1,000
instances --- 100 from each class --- where the original audio was
accurately identified and the adversarial attacks proved successful.
The attack methods used to generate AEs were the same as those listed
in FoolBox, 
as in
Experiments 1 through 3.
Given the notably low success rate of 0.3\% obtained with LS attack in
the FoolBox framework, this study focused solely on white-box attacks.
The re-attack parameters were consistent with those used in
Experiments 1 through 3, as shown in
Table~\ref{tbl:proposed_method_parameter}.

Table~\ref{tbl:defense_acc_sound} details the rectification success
rates of AEs with the proposed method, and
Table~\ref{tbl:perturb_sound} shows the perturbation amounts during
re-attacks.
The achievement of the rectification success rates of over 97\% across
all attack methods, with minimal perturbations, underscores the
robustness of the proposed method for the DNN for speech recognition.

\begin{table}[t]
\begin{center}
\caption{Experiment 4: Rectification performance in speech recognition.}\label{tbl:defense_acc_sound}
{
\begin{tabular}{@{~}c@{~}||@{~}c@{~}|@{~}c@{~~}c@{~~}c@{~~}c@{~~}c@{}}
\hline
\multirow{2}{*}{Dataset} & \multirow{2}{*}{\begin{tabular}[c]{@{}c@{}}Re-attack\\ method\end{tabular}} & \multicolumn{5}{c}{Attack method}        \\ \cline{3-7} 
& & FGSM  & BIM   & DF    & CW    & JSMA  \\ \hline \hline
Google
&FGSM                                                                     & 0.979 & 0.998 & 0.997 & 1.000 & 1.000 \\
Speech
&BIM                                                                      & 0.979 & 0.997 & 0.996 & 1.000 & 1.000 \\
Commands
&DF                                                                       & 0.979 & 0.998 & 0.998 & 1.000 & 1.000 \\ \hline 
\end{tabular}}
\end{center}
\end{table}

\begin{table}[t]
  \begin{center}
  \caption{Experiment 4: Perturbation amount of re-attacks in speech recognition.}\label{tbl:perturb_sound}
  {
  \begin{tabular}{@{~}c@{~}||@{~}c@{~}|@{~}c@{~~}c@{~~}c@{~~}c@{~~}c@{}}
  \hline
  \multirow{2}{*}{Dataset} & 
  \multirow{2}{*}{\begin{tabular}[c]{@{}c@{}}Re-attack\\ method\end{tabular}} & \multicolumn{5}{c}{Attack method}        \\ \cline{3-7} 
                                                                                        &  & FGSM  & BIM   & DF    & CW    & JSMA  \\ \hline \hline
Google   & FGSM   & 0.158 & 0.135 & 0.316 & 0.082 & 0.240 \\
Speech   & BIM    & 0.114 & 0.077 & 0.222 & 0.018 & 0.195 \\
Commands & DF     & 0.052 & 0.036 & 0.098 & 0.011 & 0.094 \\ \hline
  \end{tabular}}
  \end{center}
\end{table}

\section{Conclusion}
\label{sec:Conclusion}

This study introduces a simple yet effective method to rectify AEs by
re-attacking them to achieve the correct classification results of their
original inputs. 
The proposed method leverages AE vulnerabilities to rectify them,
enabling its application to DNNs, irrespective of the input signal
type such as images or audio.
Through a series of experiments, we successfully demonstrate that the
proposed method is more stable in rectifying AEs generated by various
attack methods than conventional ones.
Our findings highlight the effectiveness of the proposed method
against AEs generated by black-box and targeted attacks.

In the future, investigations focusing on the expansion of the
application scope of the proposed method is expected, extending it to
various modalities including language.
Furthermore, we investigate the feasibility of our proposed method as
an indicator of AE characteristics.

\appendices

\section{%
Robustness of the proposed method against parameter values
}
\label{ssec:robustness}

Because the proposed method specializes in rectifying AEs, it operates
independently of the specific settings of control parameters used
during re-attacks.
To verify this, we conducted experiments altering parameter $\epsilon$
in FGSM when using it for re-attacks within our method.

Table~\ref{tbl:tuning_cifar10} shows the rectification success rates
on CIFAR-10 when $\epsilon$ was set to 0.001, 0.01, 0.1, 1.0, and
10.0.
Compared with RS\&V, as shown in Table~\ref{tbl:rsv_comparison}, our
method maintains a high success rate even when $\epsilon$ deviates
from the default value of 1.0 in FoolBox.

\begin{table}[t]
\centering
\caption{%
Robustness of the proposed method using FGSM for control parameter $\epsilon$.
}
\label{tbl:tuning_cifar10}
\begin{tabular}{@{~}l@{~}|@{~}c@{~~}c@{~~}c@{~~}c@{~~}c@{~}|@{~}c@{~~}c@{}}
\hline

\multirow{2}{*}{$\epsilon$ in FGSM} & \multicolumn{7}{@{~}c@{}}{Attack method} \\ \cline{2-8}
\multirow{2}{*}{for re-attack}      & \multicolumn{5}{@{~}c@{~}|@{~}}{White-box} 
                                    & \multicolumn{2}{@{~}c@{}}{Black-box}    \\ \cline{2-8}
                                    & FGSM & BIM & DF & CW & JSMA & LS & HSJA \\ \hline
0.001	& 0.992	& 0.999	& 0.981	& 1.000	& 0.994	& 0.762	& 1.000 \\
0.01	& 0.992	& 1.000	& 1.000	& 1.000	& 0.994	& 0.911	& 1.000 \\
0.1	& 0.992	& 1.000	& 1.000	& 1.000	& 0.994	& 0.911	& 1.000 \\
1.0	& 0.992	& 1.000	& 1.000	& 1.000	& 0.994	& 0.911	& 1.000 \\
10.0	& 0.993	& 0.999	& 1.000	& 1.000	& 0.995	& 0.914	& 0.999 \\ \hline
\end{tabular}
\end{table}

\bibliographystyle{unsrt}
\bibliography{refs}

\begin{IEEEbiography}
[{\includegraphics[width=1in,height=1.25in,clip,keepaspectratio]{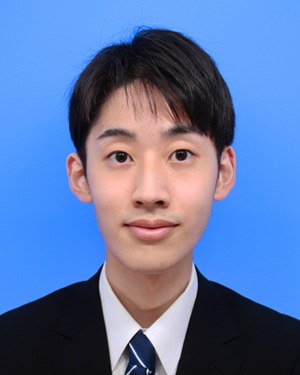}}]
{Fumiya Morimoto} received his Bachelor's degree in Engineering from
Kagoshima University, Japan in 2023. 
He is currently a master course student of Department of Engineering,
Graduate School of Science and Engineering, Kagoshima University.
His research interest includes AI security, specifically adversarial defense.
\end{IEEEbiography}

\begin{IEEEbiography}
[{\includegraphics[width=1in,height=1.25in,clip,keepaspectratio]{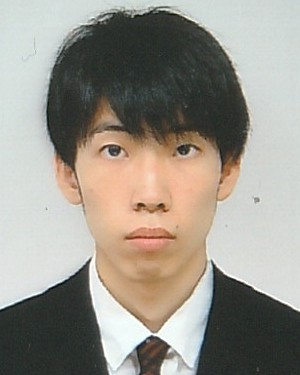}}]
{Ryuto Morita} received his Bachelor's degree in Engineering from
National Institute of Technology, Kagoshima College, Japan in 2023. 
He is currently a master course student of Department of Engineering,
Graduate School of Science and Engineering, Kagoshima University.
His research interest includes AI security, specifically adversarial defense for speech recognition.
\end{IEEEbiography}

\begin{IEEEbiography}
[{\includegraphics[width=1in,height=1.25in,clip,keepaspectratio]{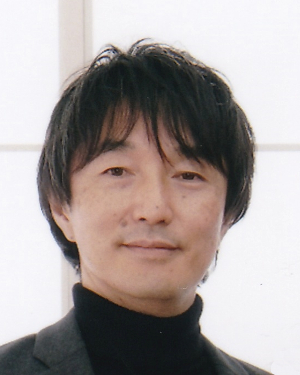}}]
{Satoshi Ono}
received his Ph.D. degree in Engineering at University of
Tsukuba in 2002. He worked as a Research Fellow of the Japanese
Society for the Promotion of Sciences (JSPS) from 2001 to
2003. Subsequently, he joined Department of Information and Computer
Science, Graduate School of Science and Engineering, Kagoshima
University as a Research Associate.  He is currently a Professor in
Department of Information Science and Biomedical Engineering in
Kagoshima University.  He received JSAI Annual Conference Award 2023,
IWAIT2020 best paper award, TAAI2019 excellent paper award, IPSJ
Yamashita SIG research award 2012, etc. He is a member of IEEE, IPSJ,
IEICE, and JSAI.  His current research focuses on evolutionary
computation, machine learning, and their applications to real world
problems.
\end{IEEEbiography}

\EOD

\end{document}